\documentclass{jfm}
\renewcommand{\figurename}{figure}
\usepackage[utf8x]{inputenc}
\usepackage{psfrag}
\usepackage{amssymb, amsmath}
\usepackage{siunitx}
\usepackage{color}

\usepackage{natbib}

\usepackage{url}

\DeclareSIUnit{\wtpercent}{wt\%}

\usepackage{upgreek}
\begin{document}

\shorttitle{Marangoni and Rayleigh convection in binary droplets} 
\shortauthor{C. Diddens et al.} 

\title{Competing Marangoni and Rayleigh convection in evaporating binary droplets}

\author{
Christian~Diddens\corresp{\email{c.diddens@utwente.nl}}\aff{1,2},
Yaxing~Li\aff{1},
Detlef~Lohse\corresp{\email{d.lohse@utwente.nl}}\aff{1,3}
}

\affiliation{
\aff{1}
Physics of Fluids group, Department of Science and Technology, Mesa+ Institute, Max Planck Center for Complex Fluid Dynamics and 
J. M. Burgers Centre for Fluid Dynamics, University of Twente, P.O. Box 217, 7500 AE Enschede, The Netherlands
\aff{2}
Department of Mechanical Engineering, Eindhoven University of Technology, P.O. Box 513, 5600 MB Eindhoven, The Netherlands
\aff{3}
Max Planck Institute for Dynamics and Self-Organization, 37077 G\"ottingen, Germany
}

\maketitle

\begin{abstract}
For a small sessile or pendant droplet it is generally assumed that gravity does not play any role once the Bond number is small. This is even assumed for evaporating binary sessile or pendant droplets, in which convective flows can be driven due to selective evaporation of one component and the resulting concentration and thus surface tension differences at the air-liquid interface. However, recent studies have shown that in such droplets gravity indeed can play a role and that natural convection can be the dominant driving mechanism for the flow inside evaporating binary droplets (Edwards et al., Phys. Rev. Lett. \textbf{121}, 184501 (2018); Li et al., Phys. Rev. Lett. \textbf{122}, 114501 (2019)). In this study, we derive and validate a quasi-stationary model for the flow inside evaporating binary sessile and pendant droplets, which successfully allows to predict the prevalence and the intriguing interaction of Rayleigh and/or Marangoni convection on the basis of a phase diagram for the flow field expressed in terms of the Rayleigh and Marangoni numbers.
\end{abstract}

\section{Introduction}
\label{sec:intro:intro}
Evaporating droplets frequently occur in nature and applications, be it a rain droplet evaporating on a leaf, a droplet on a hot surface in spray cooling, a droplet of insecticides sprayed on a leaf or an inkjet-printed ink droplet on paper. Many of such droplets are multicomponent, i.e. consisting of a mixture of liquids. From the physical point of view, an evaporating multicomponent droplet in a host gas is paradigmatic for combined multi-phase and multi-component flow including a phase transition. Scientifically, this process encompasses the various fields of fluid mechanics, thermodynamics and also aspects from the field of chemistry. The evaporation dynamics is also relevant for the deposit left behind the evaporation of a particle-laden droplet. Here, pioneering work was done by \citet{Deegan1997a} around 20 years ago, when they identified the coffee-stain effect -- i.e. the phenomenon of finding a typical ring structure of deposited particles after the evaporation of a coffee droplet -- and successfully explained it by the combination of a non-uniform evaporation rate along the droplet interface and a pinned contact line. In applications, one usually wants to prevent such coffee-stain effect, e.g. for obtaining a homogeneous deposition pattern in inkjet printing \citep{Park2006,Kuang2014,Hoath2016,Sefiane2014}. For reviews on evaporating pure droplets we refer to \citet{Cazabat2010} and \citet{Erbil2012}.

Preventing the coffee-stain effect can be achieved by altering the flow inside the droplet during the drying process by inducing gradients in the acting forces. 
Focussing on the interfacial forces first, a tangential gradient of the surface tension along the liquid-gas interface leads to the well-known Marangoni effect, i.e. a tangential traction that drives the liquid towards positions of higher surface tension \citep{Scriven1960,Pearson1958}. By that, the entire flow in the droplet can be altered from the typical outwards flow towards the contact line to a recirculating flow driven by a persistent Marangoni effect \citep{Hu2006}. For the case of a pure droplet, the necessary gradient in surface tension can be generated by thermal effects, e.g. either self-induced by latent heat of evaporation or externally imposed by heating or cooling the substrate \citep{Girard2006,Sodtke2008,Dunn2009,Tam2009}. 

The other mechanism to induce Marangoni flow is known as solutal Marangoni effect, which is usually much stronger. For solutal Marangoni flow, the droplet must consist of more than one component, e.g. a solvent and one or more surfactants \citep{Still2012,Marin2016,Kwiecinski2019} or a solvent and possibly multiple co-solvents \citep{Sefiane2003a,Christy2011,Tan2016a,Li2018} or dissolved salts \citep{Soulie2015,Marin2019}. For a recent perspective review on droplets consisting of more than one component, we refer to \citet{Lohse2020}. The difference in the volatilities of the individual constituents leads to preferential evaporation of one or the other component and thereby compositional gradients are induced. Since the surface tension is a function of the composition and due to the non-uniform evaporation profile, a surface tension gradient along the liquid-gas interface can build up and result in a similar Marangoni circulation as in the thermally-driven case. The nature of the resulting flow can be quite different, mostly depending on whether the evaporation process leads to an overall decreasing or increasing surface tension, i.e. whether the more volatile component has a higher or lower surface tension than the less volatile component.

In a binary droplet consisting e.g. of water and glycerol, with water being more volatile and having the higher surface tension, the overall surface tension decreases during the preferential evaporation of water and the resulting Marangoni flow is usually regular, axisymmetric and directed towards the position of the lowest evaporation rate of water, i.e. towards the contact line for contact angles above $\SI{90}{\degree}$ and towards the apex for contact angles below $\SI{90}{\degree}$ \citep{Diddens2017a,Diddens2017c}. 

On the contrary, e.g. in case of a binary droplet consisting of water and ethanol, where the overall surface tension increases due to the predominant evaporation of ethanol, the typical Marangoni effect is way more violent and chaotic \citep{Christy2011,Bennacer2014}. Here, in particular, the axial symmetry of the droplet is usually broken, leading to a complicated scenario of initially chaotic flow driven by the solutal Marangoni effect and followed by either thermal Marangoni flow or the typical coffee-stain flow, when almost only water is left \citep{Diddens2017}. Remarkably, the presence of a strong Marangoni effect can also have a significant influence on the shape and wetting behaviour of droplets \citep{Tsoumpas2015,Karpitschka2017}.
Finally, the evaporation of mixture droplets can show a variety of additional intriguing phenomena, e.g. multiple phase changes and microdroplet nucleation in ternary droplets like ouzo \citep{Tan2016a,Tan2017}, and phase segregation in binary droplets \citep{Li2018} or rather homogeneous deposition patterns by an interplay of Marangoni flow, surfactants and polymers \citep{Kim2016}. 

As highlighted above, besides the gradient in the surface tension, i.e. in the interfacial forces, also gradients in the mass density, i.e. in the bulk force due to gravity, can influence the flow by natural convection. Similar to the surface tension, the mass density is a function of the temperature and, in the case of mixtures, of the composition, so that thermally and solutally driven natural convection can be realized in evaporating droplets. Flow driven by natural convection is one of the most important fields of fluid mechanics, as e.g. in Rayleigh-Bénard systems, however, these are usually investigated at large spatial dimensions. 
For small droplets, on the other hand, one would naively expect that the flow in case of thermal or solutal gradients is predominantly driven by the Marangoni effect, since a small droplet is associated with a small Bond number and hence surface tension effects would dominate over gravity. As a consequence, most studies on droplet evaporation focus on the Marangoni effect, but disregard the presence of natural convection by this argument.

Recent studies, however, showed that even for small droplets with small Bond numbers, the internal flow can be decisively determined by natural convection and not by Marangoni flow \citep{Edwards2018,Li2019}. This has been even found at the later stages of water-ethanol droplets, which initially show a very intense chaotic Marangoni flow \citep{Edwards2018}. Obviously, these findings give rise to the following question: Under what circumstances which kind of flow pattern can be found in an evaporating binary droplet, i.e. when is the flow dominated by the Marangoni effect and when by natural convection?

In this manuscript, we answer this questions by carefully investigating both kinds of driving forces and their mutual interaction. The corresponding effects can be quantified by non-dimensional numbers, namely the Marangoni number for flow due to surface tension gradients and the Rayleigh (or Archimedes/Grashof) number for the natural convection. By considering quasi-stationary instants during the drying process, these numbers successfully allow to predict the flow inside the droplets on the basis of phase diagrams in the $\mbox{\it Ra}$-$\mbox{\it Ma}$ parameter space. We also validated these phase diagrams with full simulations and corresponding experiments.

The paper is organised as follows: We will first present the complete set of dynamical equations describing the evaporation of a binary mixture droplet. In section \ref{sec:motiv:motiv}, these equations are solved to discuss an illustrative example case. We will then introduce the quasi-stationary approximation in section \ref{sec:simpmodel:simpmodel} and discuss the phase diagrams obtained by this model in section \ref{sec:phasediag:phasediag}. The paper ends with a conclusion and a comparison with experimental data in the appendix. 

\section{Governing Equations}
\label{sec:detmodel:detmodel}

\begin{figure}\centering\includegraphics[width=0.99\textwidth]{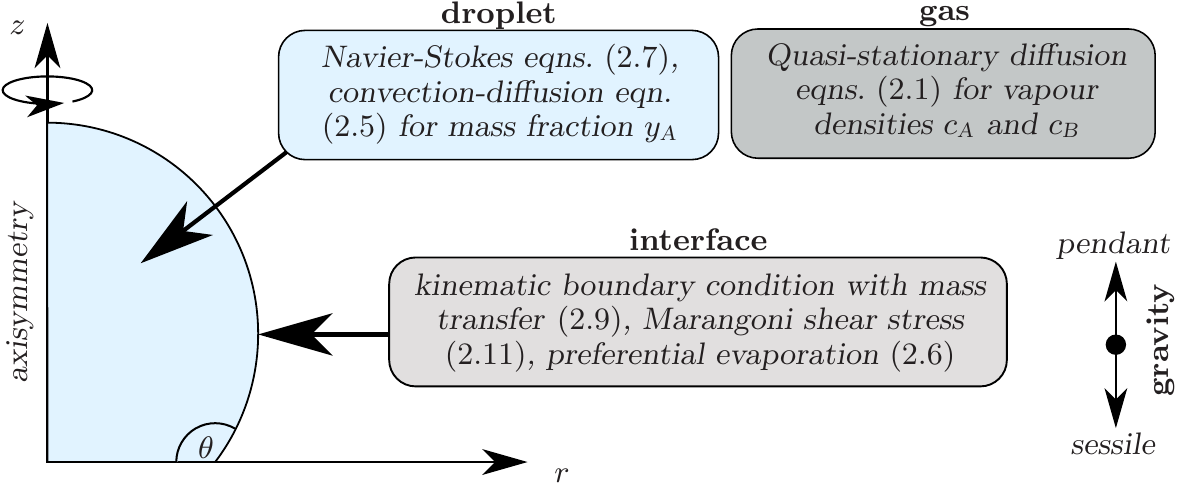}
\caption{Schematic of the model. The problem is considered to be axisymmetric and isothermal. The flow and the advection-diffusion equation for the composition in the droplet is solved with consideration of gravity and the composition-dependence of the liquid mass density, dynamic viscosity and diffusivity. The transport in the gas phase is assumed to be diffusion-limited. At the interface, Raoult's law is used to enforce the vapour-liquid equilibrium, mass transfer due to evaporation is considered and the Marangoni shear stress due to a composition-dependent surface tension is taken into account. }
\label{fig:detmodel:scheme}
\end{figure}

The evaporation of a mixture droplet is a multi-phase and free interface problem with multi-component dynamics in both the liquid and gas phase. For a binary droplet, the liquid is constituted by two components, $\alpha=\text{A},\,\text{B}$, whereas the gas phase is in general a ternary gas mixture of the ambient medium, e.g. air, and the vapours of the components $\text{A}$ and $\text{B}$.
When the droplet evaporates at a temperature $T$ far below the boiling point and in absence of forced or strong natural convection, the diffusive dynamics in the gas phase can be approximated by the quasi-stationary Laplace equation \citep{Deegan2000a,Hu2002a,Popov2005,Diddens2017a}
\begin{equation}
\nabla^2c_{\alpha}=0
\label{eq:detmodel:laplgas}
\end{equation}
for the vapour concentrations $c_{\alpha}$, i.e. the partial mass densities. The corresponding boundary conditions are given by the vapour-liquid equilibrium according to Raoult's law at the liquid-gas interface and the ambient vapour concentration at infinity, i.e.
\begin{align}
c_{\alpha}&=c_{\alpha}^{\text{eq}}=c_{\alpha}^{\text{pure}}\gamma_{\alpha}x_{\alpha} &\text{at the liquid-gas interface and} \label{eq:detmodel:raoults} \\
c_{\alpha}&=c_{\alpha}^\infty &\text{far away from the droplet}\,,
\label{eq:detmodel:laplgasbc}
\end{align}
where $c_{\alpha}^{\text{pure}}$ is the saturation vapour concentration in case of the pure liquid $\alpha$ and $x_{\alpha}$ is the liquid mole fraction. The activity coefficients $\gamma_{\alpha}$ account for thermodynamic non-idealities and are functions of the composition, i.e. of $x_{\alpha}$. Neglecting the small contribution of the Stefan flow at temperatures sufficiently below the boiling point, the evaporation rates $j_{\alpha}$ are given by the diffusive fluxes at the liquid-gas interface, i.e. 
\begin{equation}
j_{\alpha}=-D_{\alpha}^{\text{vap}}\partial_nc_{\alpha} \,.
\label{eq:detmodel:evaprate}
\end{equation}
While the dynamics in the gas phase can be considered in the diffusive and quasi-stationary limit, convection can be dominant in the liquid phase, which can be attributed to the typical diffusion coefficients, namely $D_{\alpha}^{\text{vap}}\sim \SI{e-5}{\meter^2\per\second}$ in the gas phase and $D\sim\SI{e-9}{\meter^2\per\second}$ in the liquid phase.
Therefore, the liquid phase has to be described by the full convection-diffusion equation for the liquid mass fraction $y_{\alpha}$, which is expressed for the component A only due to the identity $y_{\text{A}}+y_{\text{B}}=1$, i.e.
\begin{equation}
\rho\left(\partial_t y_{\text{A}} + \boldsymbol{u}\bcdot\bnabla y_{\text{A}} \right) = \bnabla\bcdot\left(\rho D \bnabla y_{\text{A}}\right) \,. 
\label{eq:detmodel:massfracconvA}
\end{equation}
The liquid density $\rho$ and the mutual diffusivity $D$ are in general functions of the composition, i.e. of $y_{\text{A}}$. The mass transfer rates $j_{\alpha}$ due to evaporation induce a change in composition at the liquid-gas interface. Using the mass transfer expression $j_{\alpha}=\rho y_{\alpha}(\boldsymbol{u}_\alpha-\boldsymbol{u}_{\text{I}})\bcdot\boldsymbol{n}$ with $\boldsymbol{u}_\alpha$ denoting the velocity of component $\alpha$ and $\boldsymbol{u}_{\text{I}}$ and $\boldsymbol{n}$ denoting the interface velocity and normal, respectively, this compositional change can be expressed by a Robin boundary condition for Eq. \eqref{eq:detmodel:massfracconvA} in the frame co-moving with the interface in normal direction, namely
\begin{equation}
-\rho D\bnabla{}y_{\text{A}}\bcdot \boldsymbol{n}= y_{\text{B}}j_{\text{A}}- y_{\text{A}}j_{\text{B}}= (1-y_{\text{A}})j_{\text{A}}-y_{\text{A}}j_{\text{B}}\,.
\label{eq:detmodel:massfluxbc}
\end{equation}
Finally, the flow in the droplet is given by the Navier-Stokes equations
\begin{align}
\rho\left(\partial_t\boldsymbol{u}+\boldsymbol{u}\bcdot \bnabla\boldsymbol{u}\right)&=-\bnabla p + \bnabla\bcdot\left( \mu \left(\bnabla \boldsymbol{u} + (\bnabla \boldsymbol{u})^\text{t} \right) \right) + \rho g \boldsymbol{e}_z \label{eq:detmodel:nseq} \\
\partial_t \rho + \bnabla\bcdot(\rho\boldsymbol{u})&=0\,.
\label{eq:detmodel:contieq}
\end{align}
Here, we have chosen the $z$-axis to point towards the apex of the droplet, i.e. a sessile droplet and a pendant droplet can be realized by negative and positive values for $g$, respectively. Note that the viscosity $\mu$ and the mass density $\rho$ are in general functions of the composition $y_{\text{A}}$. A dependency on the temperature is disregarded in the following due to the fact that thermal effects at lower temperatures are usually considerably inferior to the impact of solutal gradients.
For the contact line dynamics, we are focussing here on a pinned contact line, i.e. evaporation in the constant radius mode (\textsl{CR-mode}, \citet{Picknett1977,Stauber2014}). To resolve the incompatibility of a no-slip boundary condition at the substrate and the evaporative mass loss at the contact line, we impose a Navier-slip boundary condition with a small slip-length in the nanometre scale instead. This effectively resembles a no-slip boundary condition in the main part of the droplet-substrate interface, but still allows for a consistent mass transfer at the contact line. The free liquid-gas interface is subject to the kinematic boundary condition considering the mass transfer, i.e.
\begin{equation}
\left(\boldsymbol{u}-\boldsymbol{u}_{\text{I}}\right)\bcdot\boldsymbol{n}=\frac{1}{\rho}(j_{\text{A}}+j_{\text{B}})
\label{eq:detmodel:kinematicbc}
\end{equation}
and furthermore to the Laplace pressure in normal direction
\begin{equation}
-p + \mu \boldsymbol{n}\bcdot\left(\bnabla \boldsymbol{u} + (\bnabla \boldsymbol{u})^\text{t}  \right)\cdot \boldsymbol{n} = \sigma\kappa\,,
\label{eq:detmodel:laplpress}
\end{equation}
where the traction in the gas phase has been neglected due to the viscosity ratio. Here, $\sigma$ is the local surface tension, $\kappa$ the curvature of the interface and $p$ denotes the pressure difference with respect to the ambient gas pressure.
Finally, also the Marangoni shear stress in tangential direction has to be considered:
\begin{equation}
\mu \boldsymbol{n}\bcdot\left(\bnabla \boldsymbol{u} + (\bnabla \boldsymbol{u})^\text{t} \right) \cdot \boldsymbol{t} = \bnabla_{t} \sigma \bcdot \boldsymbol{t} \,.
\label{eq:detmodel:marashear}
\end{equation}
Here, $\bnabla_{t}=(\mathbf{1}-\boldsymbol{n}\boldsymbol{n})\bcdot\bnabla$ is the surface gradient operator. A sketch of the model is depicted in \figurename~\ref{fig:detmodel:scheme}.
\section{Numerical solution of the dynamical equations for an instructive example}
\label{sec:motiv:motiv}
In order to solve the given set of equations numerically, we have generalised the sharp-interface arbitrary Lagrangian-Eulerian finite element method described in \citet{Diddens2017c} by considering the gravitational force, and also validated it by a more general reimplementation of the same model with the finite element package \textsc{oomph-lib}\footnote{\url{oomph-lib.maths.man.ac.uk}} \citep{Heil2006}, which allows for interface deformations and considers the general continuity equation \eqref{eq:detmodel:contieq}. The latter method has been successfully validated against various experiments \citep{Li2018,Li2019,Gauthier2019,YanshenLi2019}.

\begin{figure}\centering\includegraphics[width=1\textwidth]{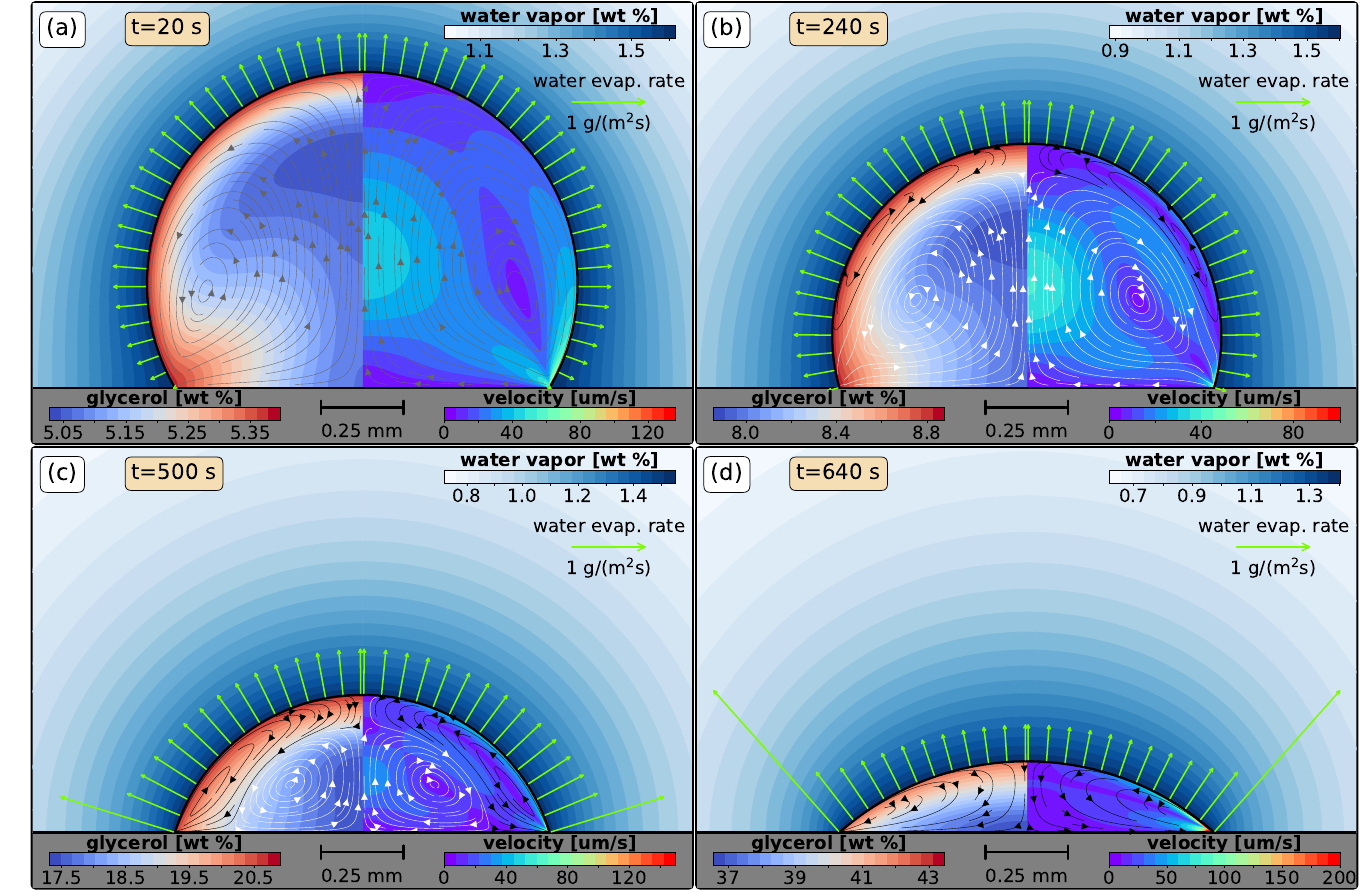}
\caption{Simulation of a \SI{1}{\micro\liter} glycerol-water droplet revealing rich flow patterns during the evaporation process. The water vapour mass fraction is shown in the gas phase, whereas the glycerol mass fraction (left) and the velocity magnitude (right) are shown inside the droplet. (a) Initially, both Rayleigh and Marangoni convection support the flow from the apex to the contact line. (b) Although the contact angle is still above $\theta>\SI{90}{\degree}$, a Marangoni-induced counter-rotating vortex (black) emerges close to the interface, whereas the bulk flow is driven by natural convection (white). (c) Due to the increased evaporation rate at the contact line for $\theta<\SI{90}{\degree}$, the Marangoni-driven vortex grows in size until (d) the vortex driven by natural convection disappears. See supplementary movie 1 for the entire simulation.}
\label{fig:motiv:example}
\end{figure}

In \figurename~\ref{fig:motiv:example}, a simulation of a sessile glycerol-water droplet (initially 5 wt.\% glycerol) with an initial volume of \SI{1}{\micro\liter} and an initial contact angle of \SI{120}{\degree} evaporating at a constant temperature of \SI{22}{\celsius} and a relative humidity of \SI{20}{\percent} is shown. The contact line remains pinned during drying and glycerol (liquid B) is assumed to be non-volatile due to its low volatility compared to water (liquid A), i.e. $c_{\text{B}}=c_{\text{B}}^\infty=0$ and $j_{\text{B}}=0$. For more details about these kind of simulations we refer to \citet{Diddens2017,Diddens2017c}, where however we did not consider of the influence of gravity. 

Initially, in \figurename~\ref{fig:motiv:example}(a), one can see a single vortex in the entire droplet, directed from the apex towards the contact line. This vortex is generated for two reasons, namely Marangoni convection and natural convection (Rayleigh convection). Due to the enhanced evaporation rate of water at the apex at the high contact angle, the water content is predominantly reduced at the top of the droplet, resulting in a lower surface tension compared to the region near the contact line.
This drives a Marangoni flow towards the contact line. Since glycerol is more dense than water, the glycerol-rich outer shell of the droplet also sinks down due to natural convection, which also results in a flow from the apex to the contact line due to the spherical geometry. Hence, both mechanisms support recirculating flow in the same direction.

Remarkably, in \figurename~\ref{fig:motiv:example}(b), the situation changes. The contact angle is still above \SI{90}{\degree}, i.e. still having the highest water evaporation rate at the top of the droplet. According to the afore-mentioned discussion, one would still naively expect the same kind of single vortex flow. However, the simulation clearly shows two vortices, one in the bulk driven by natural convection (white) and another one close to the interface, which is driven by Marangoni flow in the opposite direction (black). The reason why the Marangoni flow is reversed, i.e. why there is more water at the top of the droplet although the evaporation rate of water is still dominant at the apex, is the fact that there is enhanced water replenishment by diffusion at the apex, which compensates for the rather small difference in the evaporation rates at the top and near the contact line. This can be seen by the rather steep concentration gradient in normal direction at the apex compared to the region near the contact line. The reason of the steep concentration gradient in normal direction close at the apex is the upward directed convective water replenishment from the bulk, which is governed by the internal vortex driven by natural convection. This means that sufficiently strong natural convection in the bulk can reverse the Marangoni flow at the interface, although one would not anticipate this by just considering the profile of the evaporation rate at this contact angle.

Upon further evaporation, in \figurename~\ref{fig:motiv:example}(c), the contact angle falls below \SI{90}{\degree}, resulting in a higher water evaporation rate near the contact line. Hence, less water is present at the contact line as compared to the apex due two facts, namely the effect of preferential evaporation at the contact line and the lower replenishing diffusive flux of water from the bulk at the contact line. Thereby, the Marangoni flow gets enhanced compared to the situation in \figurename~\ref{fig:motiv:example}(b) and the relative size of the Marangoni-induced vortex at the interface grows at the expense of the counter-rotating bulk vortex by natural convection.

Finally, in \figurename~\ref{fig:motiv:example}(d), the contact angle becomes rather small so that the Marangoni flow at the interface is even stronger due to the enhanced non-uniformity of the evaporation rate. Furthermore, the influence of natural convection also diminishes rather quickly, i.e. with cubic power in terms of the length scale according to the Rayleigh number (see later for its definition), due to the reduced volume of the droplet. This results in the depicted situation, i.e. that the flow direction within the entire droplet is completely determined by the Marangoni effect.

In a nutshell, one can infer from the direct numerical simulation results in \figurename~\ref{fig:motiv:example} that there can be multiple flow scenarios during the drying of a single binary droplet, driven by an interplay of natural (i.e. Rayleigh) convection and Marangoni convection. One also clearly sees that, for a particle laden droplet, the coffee-stain effect would not occur as there is no noticeable flow towards the contact line (which for pure evaporating droplets transports the suspended particles to the rim of the pinned droplet) as compared to the strongly recirculating flow due to Marangoni flow and gravity.

\section{Quasi-stationary approximation of the dynamical equations}
\label{sec:simpmodel:simpmodel}
After discussing some possible flow scenarios by considering a representative numerical example in the previous section, we will now focus on a simplification of the full model described in section \ref{sec:detmodel:detmodel}. We generalise again from the particular case of a water-glycerol droplet to the general case of a binary droplet, where both liquids \text{A} and \text{B} are allowed to evaporate. The goal is to find the simplest model possible that allows to predict the expected flow scenario in the droplet by a minimum number of non-dimensional quantities.
 
\subsection{Evaporation Numbers}
\label{sec:simpmodel:evnos}
As shown in the example simulation, the liquid recirculates multiple times during the evaporation process due to the fast flow in the droplet. Hence, the typical liquid velocity $\boldsymbol{u}$ is much larger than the normal interface movement velocity $\boldsymbol{u}_{\text{I}}$. Moreover, this leads to a rather well-mixed droplet, i.e. with typical compositional deviations of about a few percent in terms of mass fractions. These observations allow for several simplifications of the model. First of all, the liquid composition is expanded into two terms, i.e. $y_{\text{A}}(\boldsymbol{x},t)=y_{\text{A},0}+y$, namely the spatially averaged composition $y_{\text{A},0}$, which slowly evolves over the entire drying time, and the small local composition deviations $y(\boldsymbol{x},t)$. Since the composition-dependent liquid properties are usually rather smooth functions of the composition, this separation can be transferred to a first order Taylor expansion of the liquid properties, i.e.
\begin{align}
\rho&=\rho_0+y\:\partial_{y_\text{A}}\rho\,, &\sigma&=\sigma_0+y\:\partial_{y_\text{A}}\sigma\,, \nonumber \\
\mu&=\mu_0+y\:\partial_{y_\text{A}}\mu\,, &D&=D_{0}+y\:\partial_{y_\text{A}}D\,, \label{eq:simpmodel:compoexpand}\\
c_{\text{A}}^{\text{eq}}&=c_{\text{A},0}^{\text{eq}}+y\:\partial_{y_\text{A}}c_{\text{A}}^{\text{eq}} \,, &c_{\text{B}}^{\text{eq}}&=c_{\text{B},0}^{\text{eq}}+y\:\partial_{y_\text{A}}c_{\text{B}}^{\text{eq}} \,. \nonumber 
\label{eq:simpmodel:compoexpandbla}
\end{align}
Since the averaged composition $y_{\text{A},0}$ evolves slowly, this expansion can be done at any specific time of interest during the evaporation process. In particular, this means that the coefficients of the Taylor expansions \eqref{eq:simpmodel:compoexpand} can be treated as constants during some time close to the considered instant. This allows us to introduce the following non-dimensionalized scales
\begin{equation}
\boldsymbol{x}=V^{1/3}\tilde{\boldsymbol{x}}\,,\qquad t=\frac{V^{2/3}}{D_{0}}\tilde{t}\,,\qquad \boldsymbol{u}=\frac{D_{0}}{V^{1/3}}\tilde{\boldsymbol{u}} \,,
\label{eq:simpmodel:scales}
\end{equation}
where the spatial scale is chosen in that way, that the nondimensionalized droplet volume $\tilde{V}$ becomes unity.

In a next step, the vapour fields are decomposed in a similar manner as \eqref{eq:simpmodel:compoexpand}, namely in a normalized contribution $\tilde{c}_0$ which is one at the interface and zero at infinity and a contribution $\tilde{c}_\Delta$ accounting for the effect of local composition variations on the vapour concentration via Raoult's law to the first order, i.e.
\begin{equation}
c_{\alpha}=\left(c_{\alpha,0}^{\text{eq}}-c_{\alpha}^\infty\right)\tilde{c}_0 + c_{\alpha}^\infty + (\partial_{y_\text{A}}c_{\alpha}^{\text{eq}})\tilde{c}_\Delta \,.
\label{eq:simpmodel:vapsep}
\end{equation}
The Laplace equation \eqref{eq:detmodel:laplgas} splits into two Laplace equations, i.e. $\tilde\nabla^2 \tilde{c}_0=0$ and $\tilde\nabla^2 \tilde{c}_\Delta=0$ and the boundary conditions \eqref{eq:detmodel:laplgasbc} are transformed to 
\begin{align}
\tilde{c}_0&=1\text{ and }\tilde{c}_\Delta=y &\text{at the liquid-gas interface and} \label{eq:simpmodel:raoults} \\
\tilde{c}_0&=\tilde{c}_\Delta=0  &\text{far away from the droplet}\,.
\label{eq:simpmodel:laplgasbc}
\end{align}
Thereby, the evaporation rates \eqref{eq:detmodel:evaprate} separate in the same way, i.e.
\begin{align}
j_{\alpha}=\frac{D_{\alpha}^{\text{vap}}}{V^{1/3}}\left[\left(c_{\alpha,0}^{\text{eq}}-c_{\alpha}^\infty\right)\tilde{j}_0 + (\partial_{y_\text{A}}c_{\alpha}^{\text{eq}})\tilde{j}_y \right] \,,
\label{eq:simpmodel:evaprate}
\end{align}
where $\tilde{j}_0=-\tilde\partial_n\tilde{c}_0$ only depends on the shape of the droplet, i.e. resembles the normalized evaporation profile of a homogeneous droplet, and $\tilde{j}_y=-\tilde\partial_n\tilde{c}_\Delta$ is a linear functional of $y$, i.e. the Dirichlet-to-Neumann map, accounting for deviations in the evaporation rate due to a varying interfacial composition via the composition-dependent vapour-liquid equilibrium, i.e. Raoult's law.

When dropping terms of quadratic order in $y$, the convection-diffusion equation \eqref{eq:detmodel:massfracconvA} within the droplet becomes
\begin{equation}
\partial_{\tilde t} y_{\text{A},0} + \partial_{\tilde t} y +  \tilde{\boldsymbol{u}}\bcdot\tilde\bnabla y = \tilde\nabla^2 y 
\label{eq:simpmodel:massfracconvA}
\end{equation}
and the corresponding interface boundary condition \eqref{eq:detmodel:massfluxbc} reads
\begin{equation}
-\tilde\bnabla{}y\bcdot \boldsymbol{n}= \mbox{\it Ev}_y\tilde{j}_0 +\mbox{\it Ev}_{\text{vap}}\tilde{j}_y   -   \mbox{\it Ev}_{\text{tot}} y \tilde{j}_0
\label{eq:simpmodel:massfluxbc}
\end{equation}
with the non-dimensional evaporation numbers
\begin{align}
\mbox{\it Ev}_y=&\frac{1}{\rho_0D_{0}}\Big[(1-y_{\text{A},0})D_{\text{A}}^{\text{vap}}(c_{\text{A},0}^{\text{eq}}-c_{\text{A}}^\infty)-y_{\text{A},0}D_{\text{B}}^{\text{vap}}(c_{\text{B},0}^{\text{eq}}-c_{\text{B}}^\infty)\Big] \label{eq:simpmodel:evzero}  \\
\mbox{\it Ev}_{\text{vap}}=&\frac{1}{\rho_0D_{0}}\Big[(1-y_{\text{A},0})D_{\text{A}}^{\text{vap}}\partial_{y_\text{A}}c_{\text{A}}^{\text{eq}} -y_{\text{A},0}D_{\text{B}}^{\text{vap}}\partial_{y_\text{A}}c_{\text{B}}^{\text{eq}}\Big] \label{eq:simpmodel:evy} \\
\mbox{\it Ev}_{\text{tot}}=&\frac{1}{\rho_0D_{0}}\Big[D_{\text{A}}^{\text{vap}}(c_{\text{A},0}^{\text{eq}}-c_{\text{A}}^\infty) +D_{\text{B}}^{\text{vap}}(c_{\text{B},0}^{\text{eq}}-c_{\text{B}}^\infty)\Big]\,.
\label{eq:simpmodel:evtot}
\end{align}
The number $\mbox{\it Ev}_y$ quantifies the intensity of the concentration gradient induced in the liquid by preferential evaporation of one of the components, i.e. it compares the differences of the two diffusive vapour transports in the gas phase with the mutual diffusion in the binary liquid. Since the resulting composition gradient along the interface and in the bulk is the driving mechanism for Marangoni flow and natural convection, this number will become important to quantify these processes later on. Note that dependent on the volatilities of the components and their mass fractions in the liquid, $\mbox{\it Ev}_y$ may be positive or negative.

$\mbox{\it Ev}_{\text{vap}}$ is an estimate for the influence of local variations in the liquid concentration on the preferential evaporation, i.e. the linear feedback due to the quasi-stationary diffusion in the gas phase. If the composition is rather uniform in the droplet, which is usually by fast recirculating convection, the term $\mbox{\it Ev}_{\text{vap}}\tilde{j}_y$ provides only a minor contribution in \eqref{eq:simpmodel:massfluxbc}, meaning that the profile of the evaporation rates is similar to the one of a pure droplet.  
Since $\partial_{y_\text{A}}c_{\text{A}}^{\text{eq}}>0$ and $\partial_{y_\text{A}}c_{\text{B}}^{\text{eq}}<0$, i.e. the vapour concentration of A increases and B decreases for an increasing fraction of A in the liquid, $\mbox{\it Ev}_{\text{vap}}$ is always positive. Large numbers of $\mbox{\it Ev}_{\text{vap}}$ can actually arise towards the end of the drying of a glycerol-water droplet, as discussed later on in section \ref{sec:validation:validation}.

Finally, $\mbox{\it Ev}_{\text{tot}}$ is a measure for the total evaporation speed, i.e. for the typical interface speed $\tilde{\boldsymbol{u}}_{\text{I}}$ and the volume evolution. Note that the total evaporation speed and volume evolution is a measure for the flow towards a pinned contact line, i.e. the flow leading to the coffee-stain effect. If none of the components condensates, i.e. both either evaporate or are non-volatile, the modulus of $\mbox{\it Ev}_y$ is smaller than $\mbox{\it Ev}_{\text{tot}}$. Nevertheless, since the deviation from the average composition is small, i.e. $y\ll 1$, and the Marangoni convection and/or natural convection are sufficiently large, the contribution of the latter to the flow can still be dominant compared to $\mbox{\it Ev}_{\text{tot}} y \tilde{j}_0$ in \eqref{eq:simpmodel:massfluxbc}.

\subsection{Nondimensionalized flow}
\label{sec:simpmodel:nseq}
For the Navier-Stokes equations, we employ the established Boussinesq approximation, which is valid as long as $y\partial_{y_\text{A}}\rho$ is small compared to $\rho_0$ \citep{Gray1976}.
Due to the usually small composition gradients, this assumption is valid here. Therefore, except for the body force term $\rho g \boldsymbol{e}_z$, only the zeroth order terms proportional to $\rho_0$ are kept, whereas $y\partial_{y_\text{A}}\rho$-terms are disregarded. With the same argument, terms proportional to $y\partial_{y_\text{A}}\mu$ and $y\partial_{y_\text{A}}\sigma$ can be disregarded, whenever there is a dominant term proportional to $\mu_0$ and $\sigma_0$, respectively. Following this argument, the Navier-Stokes equations can be written as  
\begin{align}
\mbox{\it Sc}^{-1}\left(\partial_{\tilde t}\tilde{\boldsymbol{u}}+\tilde{\boldsymbol{u}}\bcdot \tilde\bnabla\tilde{\boldsymbol{u}}\right)&=-\tilde\bnabla \tilde p + \tilde\bnabla\bcdot \left(\tilde\bnabla \tilde{\boldsymbol{u}} + (\tilde\bnabla \tilde{\boldsymbol{u}})^\text{t} \right)  + \mbox{\it Ra}^* y \boldsymbol{e}_z \label{eq:simpmodel:nseq} \\
\tilde\bnabla\bcdot\tilde{\boldsymbol{u}}&=0\,.
\label{eq:simpmodel:contieq}
\end{align}
Here, the shifted nondimensionalized pressure, the Schmidt number and the incomplete Rayleigh number read
\begin{equation}
\tilde p=\frac{V^{2/3}}{D_{0}\mu_0}\left(p-\rho g z\right) \,, \qquad \mbox{\it Sc}=\frac{\mu_0}{D_{0}\rho_0}\,, \qquad \mbox{\it Ra}^*=\frac{V g\partial_{y_\text{A}}\rho}{D_{0}\mu_0}\,.
\label{eq:simpmodel:reynra}
\end{equation}
The Schmidt number for liquids is usually $\mbox{\it Sc}> \num{e3}$ which suggests that the lhs of \eqref{eq:simpmodel:nseq} can be disregarded. However, since the chosen velocity and time scale in \eqref{eq:simpmodel:scales} does not necessarily coincide with the actual present scales, this argument is only valid for small Reynolds numbers. In small droplets with rather low volatilities and regular Marangoni flow, however, this assumption is surely met, e.g. $\mbox{\it Re} < 0.05$ in the case of the simulation in \figurename~\ref{fig:motiv:example}. The incomplete Rayleigh number $\mbox{\it Ra}^*$ deviates from the conventional definition of the Rayleigh number just by the lack of an estimate for the composition difference, i.e. a term like $\Delta y_{}$. The dynamic boundary conditions at the interface, \eqref{eq:detmodel:laplpress} and \eqref{eq:detmodel:marashear}, read in the Boussinesq approximation
\begin{equation}
-\tilde p + \boldsymbol{n}\bcdot\left(\tilde\bnabla \tilde{\boldsymbol{u}} + (\tilde\bnabla \tilde{\boldsymbol{u}})^\text{t}  \right)\cdot \boldsymbol{n} = \frac{1}{\mbox{\it Ca}^*}\left(\tilde\kappa+\mbox{\it Bo}\: \tilde z\right)
\label{eq:simpmodel:laplpress}
\end{equation}
\begin{equation}
\boldsymbol{n}\bcdot\left(\tilde\bnabla \tilde{\boldsymbol{u}} + (\tilde\bnabla \tilde{\boldsymbol{u}})^\text{t} \right) \cdot \boldsymbol{t} = \mbox{\it Ma}^* \tilde\bnabla_{t} y \bcdot \boldsymbol{t} \,.
\label{eq:simpmodel:marashear}
\end{equation}
Here, the non-dimensional number $\mbox{\it Ca}^*$, the Bond number and the incomplete Marangoni number read
\begin{equation}
\mbox{\it Ca}^*=\frac{D_{0}\mu_0}{V^{1/3}\sigma_0}\,,  \qquad \mbox{\it Bo}=\frac{\rho_0 g V^{2/3}}{\sigma_0}\,, \qquad  \mbox{\it Ma}^*=\frac{V^{1/3}\partial_{y_\text{A}}\sigma}{D_{0}\mu_0}\,.
\label{eq:simpmodel:mastar}
\end{equation}
Note that the definition of $\mbox{\it Ca}^*$ does not coincide with the capillary number, i.e. it does not consider the actually present typical velocity scale, i.e. the intensity of the capillary shape relaxations during evaporation cannot be inferred from $\mbox{\it Ca}^*$. However, both the real capillary number $\mbox{\it Ca}=\mu_0U/\sigma_0$ and $\mbox{\it Ca}^*$ are small in the systems considered here ($\mbox{\it Ca}<\num{1e-6}$ and $\mbox{\it Ca}^*<\num{1e-7}$ in the simulation depicted in \figurename~\ref{fig:motiv:example}). 
Similar to $\mbox{\it Ra}^*$, the incomplete Marangoni number $\mbox{\it Ma}^*$ lacks in an estimate for the composition difference, i.e. $\Delta y$, as compared to the conventional definition.
Finally, the kinematic boundary condition \eqref{eq:detmodel:kinematicbc} becomes
\begin{equation}
\left(\tilde{\boldsymbol{u}}-\tilde{\boldsymbol{u}_{\text{I}}}\right)\bcdot\boldsymbol{n}=\mbox{\it Ev}_{\text{tot}} \tilde{j}_0 + \mbox{\it Ev}_{\text{tot,vap}} \tilde{j}_y \,,
\label{eq:simpmodel:kinematicbc}
\end{equation}
where
\begin{equation}
\mbox{\it Ev}_{\text{tot,vap}}=\frac{1}{\rho_0D_{0}}\Big[D_{\text{A}}^{\text{vap}}\partial_{y_\text{A}}c_{\text{A}}^{\text{eq}} +D_{\text{B}}^{\text{vap}}\partial_{y_\text{A}}c_{\text{B}}^{\text{eq}}\Big]
\label{eq:simpmodel:evtoty}
\end{equation}
is the analogue of $\mbox{\it Ev}_{\text{vap}}$ for the total evaporation rate, i.e. the effect of a change in the saturation pressure due to a locally deviating composition on the total evaporation rate.

\subsection{Estimation of the outwards flow}
\label{sec:simpmodel:coffstain}
\begin{figure}\centering\includegraphics[width=1\textwidth]{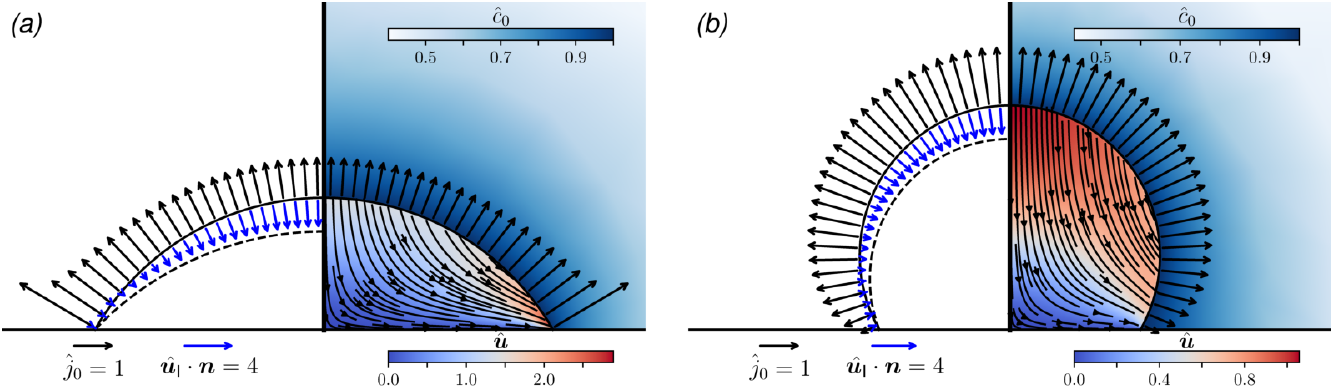}
\caption{Bulk flow for small capillary number $\mbox{\it Ca}$ and Bond number $\mbox{\it Bo}$ in absence of Marangoni flow and natural convection for a total evaporation number $\mbox{\it Ev}_{\text{tot}}=1$ and a contact angle of (a) $\theta=\SI{60}{\degree}$ and (b) $\theta=\SI{120}{\degree}$. The evaporation rate $\mbox{\it Ev}_{\text{tot}}\tilde{j}_0$ causes a volume loss, which prescribes the normal interface movement $\tilde{\boldsymbol{u}}_{\text{I}}\bcdot\boldsymbol{n}$ (depicted on the left sides). The bulk flow (right sides) is governed by Stokes flow with the normal boundary condition $\tilde{\boldsymbol{u}}\bcdot\boldsymbol{n}=\tilde{\boldsymbol{u}_{\text{I}}}\bcdot\boldsymbol{n}+\mbox{\it Ev}_{\text{tot}}\tilde{j}_0$. Apparently, the typical purely capillary-driven velocity is of order $\mbox{\it Ev}_{\text{tot}}$.  }
\label{fig:simpmodel:coffstain}
\end{figure}

Before focusing on natural convection and Marangoni flow in the droplet, it is beneficial to obtain an estimate for the velocity in the droplet in absence of these mechanisms, i.e. $\mbox{\it Ma}^*=\mbox{\it Ra}^*=0$. This case, exemplified e.g. by a pure isothermal droplet, combined with a pinned contact line represents the purely capillary-driven outward flow, which causes the coffee-stain effect. 

For small droplets, the capillary number $\mbox{\it Ca}$ is small, so that the surface tension forces according to \eqref{eq:simpmodel:laplpress} lead to an intense relaxing traction, whenever the droplet deviates from the equilibrium shape. Since the Bond number $\mbox{\it Bo}$ is small as well for small droplets, the hydrostatic term in \eqref{eq:simpmodel:laplpress} can be neglected, leading to a spherical cap with a homogeneous curvature $\tilde\kappa$ as equilibrium shape. Hence, the shape evolution and thereby the interface velocity $\tilde{\boldsymbol{u}_{\text{I}}}$ is solely given by the evaporation rate and the contact line kinetics, which is assumed to be pinned here. Since the term $\mbox{\it Ev}_{\text{tot,vap}} \tilde{j}_y$ in \eqref{eq:simpmodel:kinematicbc} is proportional to $y$, it can be disregarded with respect to $\mbox{\it Ev}_{\text{tot}} \tilde{j}_0$ in accordance with the Boussinesq approximation. As a consequence, one ends up with a linear Stokes flow problem, where the entire bulk velocity is given by the instantaneous shape relaxation, which is proportional to the rate of evaporation, i.e. to $\mbox{\it Ev}_{\text{tot}}$.

By integrating the evaporation rate $\mbox{\it Ev}_{\text{tot}} \tilde{j}_0$ one obtains the volume loss and thereby one can reconstruct the normal velocity of the interface $\tilde{\boldsymbol{u}}_{\text{I}}\bcdot\boldsymbol{n}$. The flow in the bulk $\tilde{\boldsymbol{u}}$ is subsequently given by solving the Stokes flow with the normal boundary condition $\tilde{\boldsymbol{u}}\bcdot\boldsymbol{n}=\tilde{\boldsymbol{u}_{\text{I}}}\bcdot\boldsymbol{n}+\mbox{\it Ev}_{\text{tot}}\tilde{j}_0$. In \figurename~\ref{fig:simpmodel:coffstain}, representative solutions for the bulk flow $\tilde{\boldsymbol{u}}$ with unity evaporation number, i.e. $\mbox{\it Ev}_{\text{tot}}=1$, are depicted. It is apparent that the typical bulk velocity is of the order unity, i.e. $\|\tilde{\boldsymbol{u}}\|\sim 1$. Since the flow is proportional to $\mbox{\it Ev}_{\text{tot}}$, the typical velocity corresponding to an arbitrary evaporation number $\mbox{\it Ev}_{\text{tot}}$ is hence $\|\tilde{\boldsymbol{u}}\|\sim\mbox{\it Ev}_{\text{tot}}$. This holds also for the typical interface movement, i.e. $\|\tilde{\boldsymbol{u}}_{\text{I}}\|\sim\mbox{\it Ev}_{\text{tot}}$.

\subsection{Quasi-stationary limit}
Knowing the fact that the capillary flow due to the volume loss is on the order of $\mbox{\it Ev}_{\text{tot}}$, we now focus on the contributions to the flow by Marangoni forces and natural convection. In a first step, one can consider the case where $\mbox{\it Ev}_{\text{tot}}=0$, i.e. no total mass transfer and hence a constant volume and shape of the droplet. This scenario can be realized by tuning the ambient humidities of A and B so that the evaporative mass loss of component A is balanced by the condensation of component B. In this case $\mbox{\it Ev}_{\text{tot}}=0$ and $\mbox{\it Ev}_y>0$ holds. 
Again, due to the small capillary number and the small Bond number, one can assume a spherical cap shape with volume $\tilde{V}=1$ and contact angle $\theta$, which are both constant now. Furthermore, there is no interface movement, $\tilde{\boldsymbol{u}}_{\text{I}}\bcdot\boldsymbol{n}=0$, and no total mass transfer, $\tilde{\boldsymbol{u}}\bcdot\boldsymbol{n}=0$.

By averaging \eqref{eq:simpmodel:massfracconvA} over the droplet volume $\tilde{V}=1$, defining the integrated evaporation rate 
\begin{equation}
\tilde J=\int \tilde{j}_0 \mathrm{d}\tilde A
\label{eq:simpmodel:intevap}
\end{equation}
and considering only the zeroth order term in the boundary condition \eqref{eq:simpmodel:massfluxbc} in accordance with the Boussinesq approximation, one can separate the average composition $y_{\text{A},0}$ and the deviation $y$ as follows:
\begin{align}
\partial_{\tilde t} y_{\text{A},0} &= -\mbox{\it Ev}_y \tilde J_0 \label{eq:simpmodel:avgevo} \\
\partial_{\tilde t} y +  \tilde{\boldsymbol{u}}\bcdot\tilde\bnabla y &= \tilde\nabla^2 y +\mbox{\it Ev}_y \tilde J_0 \,.
\label{eq:simpmodel:advdiffuwithcorrection}
\end{align}
Here, the term $\mbox{\it Ev}_y \tilde J_0$ assures that $y_{\text{A},0}$ is indeed the average composition and that the average of $y$ remains zero, i.e. the term compensates for the imposed composition gradient at the liquid-gas interface. As already stated in section \ref{sec:simpmodel:evnos}, this splitting holds only for limited time, since a variation in $y_{\text{A},0}$ leads to a change in the liquid properties which where used for the nondimensionalization. Usually, however, the coupled dynamics of flow $\tilde{\boldsymbol{u}}$ and compositional differences $y$ due to Marangoni and natural convection is considerable faster than $\mbox{\it Ev}_y \tilde J_0$, This was already apparent from the simulations depicted in \figurename~\ref{fig:motiv:example} and it will be validated later on in section \ref{sec:validation:validation}. Furthermore, this observation allows to focus on stationary solutions. Finally, upon introducing 
\begin{align}
\xi =\frac{y}{\mbox{\it Ev}_y}\,,
\label{eq:simpmodel:tildey}
\end{align}
one ends up at the following set of coupled equations:
\begin{align}
\tilde{\boldsymbol{u}}\bcdot\tilde\bnabla \xi &= \tilde\nabla^2 \xi +\tilde J_0 \label{eq:simpmodel:cdessnd} \\
-\tilde\bnabla \tilde p + \tilde\bnabla\bcdot \left(\tilde\bnabla \tilde{\boldsymbol{u}} + (\tilde\bnabla \tilde{\boldsymbol{u}})^\text{t} \right)  + \mbox{\it Ra}\, \xi \boldsymbol{e}_z&=0 \label{eq:simpmodel:nseqStokes} \\
\tilde\bnabla\bcdot\tilde{\boldsymbol{u}}&=0
\label{eq:simpmodel:divfree}
\end{align}
subject to the following boundary conditions 
\begin{align}
-\tilde\bnabla{}\xi\bcdot \boldsymbol{n}&= \tilde{j}_0  \label{eq:simpmodel:simpMl}\\
\tilde{\boldsymbol{u}}\bcdot\boldsymbol{n}&=0\label{eq:simpmodel:nooutfluxss}\\
\boldsymbol{n}\bcdot\left(\tilde\bnabla \tilde{\boldsymbol{u}} + (\tilde\bnabla \tilde{\boldsymbol{u}})^\text{t} \right) \cdot \boldsymbol{t} &= \mbox{\it Ma}\: \tilde\bnabla_{t} \xi \bcdot \boldsymbol{t} 
\label{eq:simpmodel:marashearss}
\end{align}
at the liquid-gas interface and 
\begin{align}
\tilde\bnabla{}\xi\bcdot \boldsymbol{n}&= 0  \label{eq:simpmodel:nopene}\\
\boldsymbol{u}&=0
\label{eq:simpmodel:noslipss}
\end{align}
at the liquid-substrate interface. Note that the simplified kinematic boundary condition \eqref{eq:simpmodel:nooutfluxss} is now compatible with the no-slip boundary condition \eqref{eq:simpmodel:noslipss} at the contact line, i.e. a slip length is not required.
Besides the contact angle $\theta$, only two parameters enter the system, namely the Marangoni number and the Rayleigh number, which read
\begin{equation}
\mbox{\it Ma}=\mbox{\it Ma}^*\mbox{\it Ev}_y=\frac{V^{1/3}\partial_{y_\text{A}}\sigma}{D_{0}\mu_0}\mbox{\it Ev}_y\,\,,\qquad \mbox{\it Ra}=\mbox{\it Ra}^*\mbox{\it Ev}_y=\frac{V g\partial_{y_\text{A}}\rho}{D_{0}\mu_0}\mbox{\it Ev}_y\,.
\label{eq:simpmodel:MaNoRaNo}
\end{equation}
Note that the characteristic numbers for both mechanisms are proportional to the induced composition gradient due to mass transfer, i.e. $\mbox{\it Ev}_y$. Of course, in particular the tangential gradient along the interface is also strongly dependent on the contact angle $\theta$, since this determines the profile of the evaporation rate $\tilde{j}_0$.
 
These equations are not only valid for the specific assumed case of $\mbox{\it Ev}_{\text{tot}}=0$, but also when the combination of Marangoni and Rayleigh flow $\tilde{\boldsymbol{u}}$ predicted by this model is considerably faster than the capillary flow, i.e. a flow situation with outwards flow leading to the coffee-stain effect. According to the estimations in section \ref{sec:simpmodel:coffstain}, this is the case if $\tilde{\boldsymbol{u}}\gg \mbox{\it Ev}_{\text{tot}}$.

\section{Phase diagram for combined Marangoni and Rayleigh convection}
\label{sec:phasediag:phasediag}

\subsection{Procedure}
Unfortunately, the analytical treatment of the model equations \eqref{eq:simpmodel:cdessnd}-\eqref{eq:simpmodel:divfree} is hampered by the geometry, which demands rather complicated toroidal coordinates, and by the very strong nonlinear coupling of $\tilde{\boldsymbol{u}}$ and $\xi$ due to the advection term. Therefore we investigate the system by numerical means. Our analysis is limited to axisymmetric solutions and we only consider the case $\mbox{\it Ev}_{\text{vap}}=0$, i.e. neglecting the feedback of the altered gas composition due to the liquid-vapour equilibrium on the local evaporation rate.
Finally, we will focus on the case $\mbox{\it Ma}\geq 0$, for which evaporation leads to an overall reduction of the surface tension, as in the case of the water-glycerol depicted in \figurename~\ref{fig:motiv:example}. This results in a regular flow, i.e. no chaotic behaviour can be found, at least not for moderate flow conditions. In the case of negative Marangoni numbers, chaotic flow patterns cannot be excluded due to the Marangoni instability \citep{Christy2011,Bennacer2014,Machrafi2010}. Of course, this spatio-temporal evolving type of flow cannot be captured within the assumption of a quasi-stationary process. One can, on the other hand, test the linear stability of the quasi-stationary solutions in the case of negative Marangoni numbers to find the transition to chaotic flow, but since also the axial symmetry is usually broken in case of negative Marangoni numbers, one also would have to generalize the entire solution procedure from axisymmetric cylindrical coordinates to the full three-dimensional problem, as done by \citet{Diddens2017}.

In order to find solutions of the system, we employed a finite element method on an axisymmetric mesh with triangular elements. We used linear basis functions for $\xi$ and $\tilde p$ and quadratic basis functions for $\tilde{\boldsymbol{u}}$, i.e. typical Taylor-Hood elements. The equations have been implemented in both \textsc{FEniCS}\footnote{\url{https://fenicsproject.org/}} \citep{LoggMardalEtAl2012a} and \textsc{oomph-lib} for mutual validation. The condition of zero velocity in normal direction, i.e. Eq. \eqref{eq:simpmodel:nooutfluxss}, has been implemented by Lagrange multipliers. For an enhanced stability in the Newton method during the solution process, it has been found beneficial to replace $\tilde J_0$ in \eqref{eq:simpmodel:cdessnd} by a Lagrange multiplier which ensures that the average of $\xi$ is zero. This removes the null space with respect to a constant shift in $\xi$ and a corresponding adjustment of the pressure $\tilde p$.

Due to the nonlinear advection term, it is in general possible that multiple solutions exist for a given parameter combination $(\theta,\mbox{\it Ma},\mbox{\it Ra})$. For the parameter ranges considered in the following, however, we are confident that we found the generic solutions due to the following strategy: For every considered contact angle $\theta$, we performed adiabatic scans along $\mbox{\it Ra}$ in increasing and decreasing direction for fixed $\mbox{\it Ma}$ and vice versa. During that, no hysteresis, i.e. bistable regions, have been found. Furthermore, by tracing these parameter paths with continuation, we have not detected any unstable branches. This has been furthermore validated by investigating the eigenvalues with a shift-inverted Arnoldi method\footnote{using \textit{Spectra} \url{https://spectralib.org}}. Finally, for each parameter combination, we performed temporal integrations of the unsteady model equations starting from a homogeneous state $\xi=0$. Since these runs converged to the same solutions as obtained by the steady parameter scans, we are sure that all solutions discussed in the following are indeed generic and stable. Note, however, that this is in general not true outside the considered parameter ranges.

\subsection{Phase diagrams}
\begin{figure}\centering\includegraphics[width=1\textwidth]{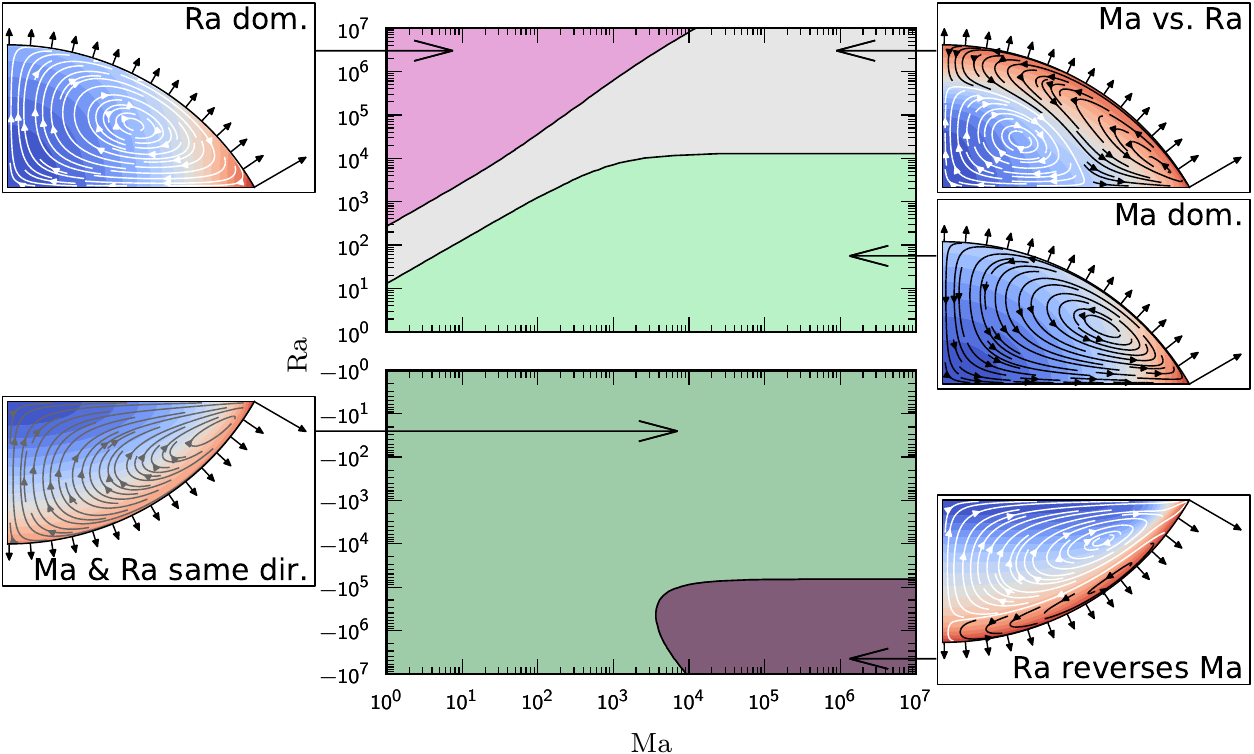}
\caption{Qualitative flow types as function of the Marangoni number $\mbox{\it Ma}$ and the Rayleigh number $\mbox{\it Ra}$ for a small contact angle $\theta=60^\circ$. For sessile droplets ($\mbox{\it Ra}>0$) with large Marangoni numbers and small Rayleigh numbers, Marangoni flow dominates in the droplet and results in a circulating flow from the contact line along the free interface towards the apex (\textsl{Ma dominant}, black streamlines). On the contrary, if $\mbox{\it Ma}$ is small and  $\mbox{\it Ra}$ is large, gravity-driven flow dominates with a flow direction from the apex along the free interface to the contact line (\textsl{Ra dominant}, white streamlines). While the other mechanism, i.e. natural convection in the regime \textsl{Ma dominant} and Marangoni flow in the regime \textsl{Ra dominant}, can still quantitatively influence the flow, only a single vortex can be found which is driven by the dominant mechanism. In between these regions, however, there is a regime where the bulk flow is driven by natural convection whereas the flow close to the interface is dominated by the Marangoni effect (\textsl{Ma vs. Ra}). Here, two counter-rotating vortices can be seen.
For pendant droplets ($\mbox{\it Ra}<0$), both mechanisms driving the flow in the same direction (\textsl{Ma \& Ra same dir.}). Hence, one cannot identify the main driving mechanism from the direction of the flow, so that the streamlines are coloured grey. If both mechanisms are sufficiently strong, however, the bulk flow due to natural convection can become so intense that the composition gradient along the interface changes direction and a flow reversal due to the Marangoni effect can arise in the vicinity of the interface (\textsl{Ra reverses Ma}).}
\label{fig:phasediag:MaRaLowCa}
\end{figure}

\begin{figure}\centering\includegraphics[width=1\textwidth]{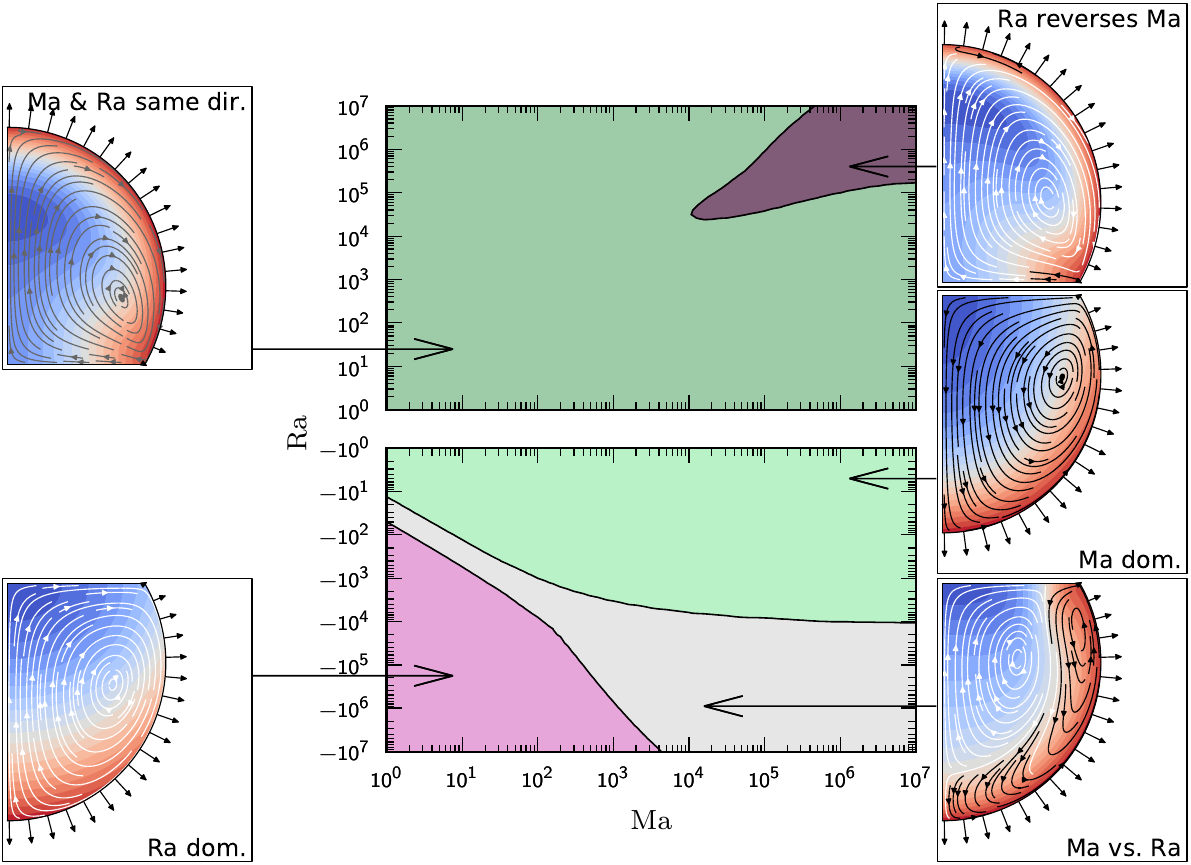}
\caption{Qualitative flow types as function of the Marangoni number $\mbox{\it Ma}$ and the Rayleigh number $\mbox{\it Ra}$ for a high contact angle $\theta=120^\circ$.
Since the direction of the Marangoni flow is reversed in comparison to the case  $\theta<90^\circ$ (cf. \figurename~\ref{fig:phasediag:MaRaLowCa}), the diagram qualitatively flips upside down. Now, for sessile droplets ($\mbox{\it Ra}>0$) both mechanisms act in the same direction (\textsl{Ma \& Ra same dir.}) and for sufficiently intense driving, the natural convection in the bulk can reverse the composition gradient at the interface, leading to a Marangoni-driven reversal close to the interface (\textsl{Ra reverses Ma}). For pendant droplets ($\mbox{\it Ra}<0$), either Marangoni flow or natural convection dominates (\textsl{Ma dominant}/\textsl{Ra dominant}), or the bulk flow is driven by natural convection, whereas the interfacial flow is governed by Marangoni flow (\textsl{Ma vs. Ra}).
}
\label{fig:phasediag:MaRaHighCa}
\end{figure}

The phase diagrams for small and large contact angles, i.e. for $\theta=\SI{60}{\degree}$ and $\theta=\SI{120}{\degree}$, are depicted in figures \figurename~\ref{fig:phasediag:MaRaLowCa} and \figurename~\ref{fig:phasediag:MaRaHighCa}, respectively. Here we have assumed, as in the case of the glycerol-water droplet, that the blue liquid A (e.g. water) is more volatile, less dense but associated with a higher surface tension than the red liquid B (e.g. glycerol). This means by definition that $\mbox{\it Ma}>0$ holds and that a sessile droplet is described by $\mbox{\it Ra}>0$ whereas a pendant droplet is given by $\mbox{\it Ra}<0$ .

Depending on the Marangoni number $\mbox{\it Ma}$, the Rayleigh number $\mbox{\it Ra}$ and the contact angle $\theta$, different qualitative flow scenarios can be found. For high Marangoni numbers and small Rayleigh numbers, the Marangoni flow dominates (\textsl{Ma dominant}) and vice versa (\textsl{Ra dominant}). In between, however, for sessile droplets with a contact angle below $\SI{90}{\degree}$ and for pendant droplets with a contact angle above $\SI{90}{\degree}$, there is a region where the Marangoni effect determines the flow direction at the interface, whereas the bulk flow is driven by natural convection (\textsl{Ma vs. Ra}). In the opposite cases, i.e. for pendant droplets with $\theta<\SI{90}{\degree}$ and sessile droplet with $\theta>\SI{90}{\degree}$, both mechanisms drive a flow in the same direction, so that one cannot directly distinguish between the two mechanisms driving the flow (\textsl{Ma \& Ra same dir.}). In the limit of very strong driving of both mechanisms, however, natural convection can become so intense, that the surface tension gradient is reversed, leading to a Marangoni-induced reversal of the flow at the interface (\textsl{Ra reverses Ma}). This effect can be explained by the distortion of the internal composition field due to natural convection. For pendant droplets with $\theta<\SI{90}{\degree}$, the composition gradient in the bulk in normal direction is much more pronounced near the contact line as opposed to the apex. As a consequence, the diffusive replenishment of the blue liquid at the interface is enhanced near the contact line so that in fact more blue liquid, i.e. the one with higher surface tension, can be found near the contact line instead of at the apex -- despite of its higher volatility and the pronounced evaporation rate at the contact line. The resulting Marangoni flow is therefore reversed as anticipated by considering the profile of the evaporation rate alone. The same explanation holds for the case $\theta>\SI{90}{\degree}$, except that one finds a more pronounced normal composition gradient near the apex as compared to the region close to the contact line, and the situation is reversed.

All transitions between the afore-mentioned regimes are continuous. The drawn phase boundaries are defined by the emergence or disappearance of a second vortex. There is no bifurcation and/or hysteresis present at the boundaries of the regimes. In supplementary movie 2, a path through the parameter space is traversed and the corresponding stationary solution is shown, which illustrates the behaviour of the flow upon crossing the phase region boundaries, i.e. how the stationary solution gradually changes between single and two-vortex solutions.

\begin{figure}\centering\includegraphics[width=1\textwidth]{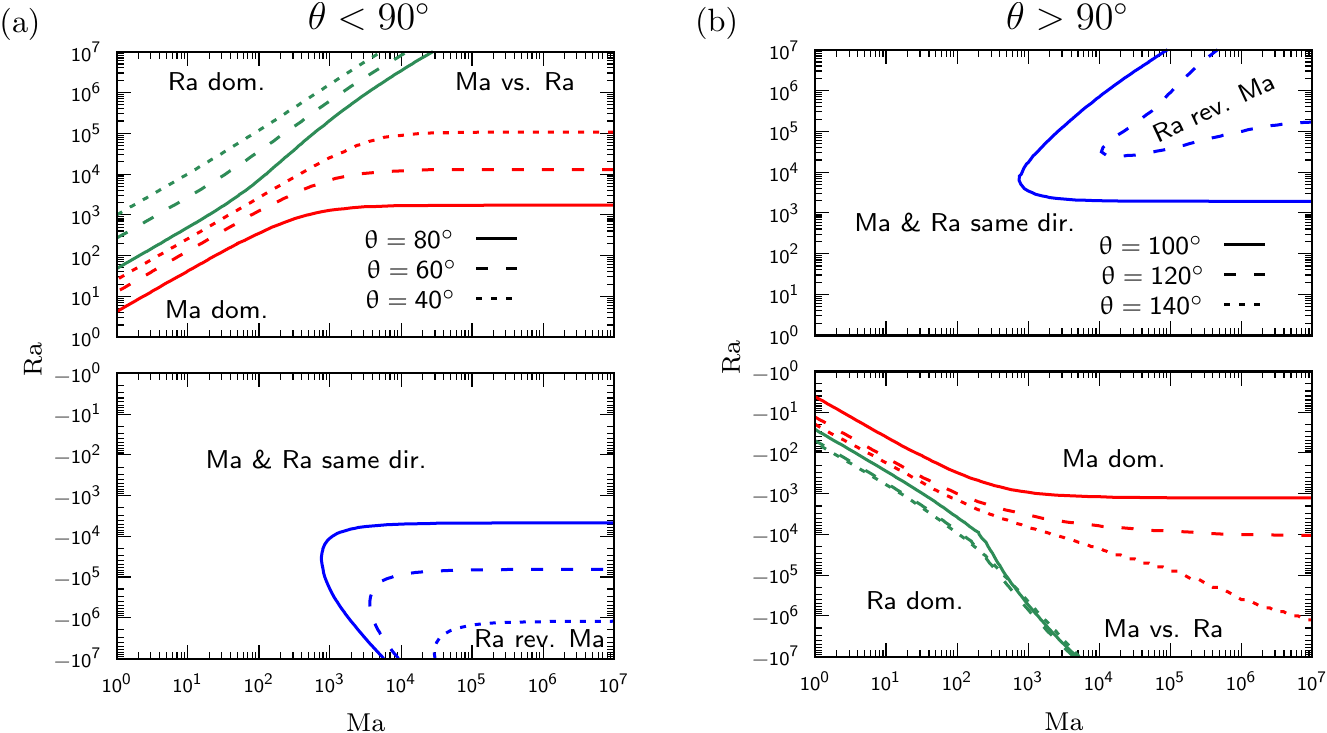}
\caption{Influence of the contact angle $\theta$ on the boundaries of the phase diagram for (a) $\theta<\SI{90}{\degree}$ and (b) $\theta>\SI{90}{\degree}$.}
\label{fig:phasediag:CaDep}
\end{figure}

Finally, we also investigate the contact angle dependence of the phase diagrams by showing the corresponding regions for $\theta=\SI{40}{\degree},\,\SI{60}{\degree}$ and $\SI{80}{\degree}$ in \figurename~\ref{fig:phasediag:CaDep}(a) and for $\theta=\SI{100}{\degree},\,\SI{120}{\degree}$ and $\SI{140}{\degree}$ in \figurename~\ref{fig:phasediag:CaDep}(b). Obviously, the phase boundaries are shifted, but qualitative differences in the phase diagrams cannot be found.

Note again that we have assumed in the phase diagrams that the more volatile liquid (blue) is less dense in the insets in the phase diagrams. In the other case, the droplets depicted in the insets are required to be mirrored vertically, as the Rayleigh number is then negative for sessile droplets and positive for pendant droplets.
Furthermore, it is noteworthy that these diagrams are for pinned droplets (CR-mode) and droplets with a constant contact angle (CA-mode), as only stationary solutions are considered anyhow. As long as the dominant velocity contribution is given by recirculating flow due to Marangoni and/or natural convection, any capillary flow due to shape relaxations can be disregarded.  Finally, the diagrams can also be used for condensation instead of evaporation, as long as the $\mbox{\it Ma}>0$ holds. For condensation of component A, $\mbox{\it Ev}_y<0$ holds, so that $\mbox{\it Ma}>0$ is true if component A has the lower surface tension, i.e. $\partial_{y_\text{A}}\sigma<0$. We therefore anticipate that the diagrams can predict the flow when ethanol condenses on a pure water droplet, whereas it would fail to predict the flow when water condenses on a pure glycerol droplet ($\mbox{\it Ma}<0$). In fact, the latter case has been investigated experimentally, showing indeed chaotic cellular flow structures \citep{Shin2016}. 
\section{Validation of the quasi-stationary approximation against the full numerical simulation}
\label{sec:validation:validation}
Since there were a number of assumptions made in the simplification of the problem, it is necessary to validate the predicted flow by comparing it with results of the full numerical simulation, i.e. with the full set of equations as described in section \ref{sec:detmodel:detmodel}. We focus on the representative simulation depicted in \figurename~\ref{fig:motiv:example}. At each instant in time, we have extracted the spatially averaged water mass fraction $y_{\text{A},0}$, the volume $V$ to determine the spatial scale $\sqrt[3]{V}$ and the contact angle $\theta$ from the simulation. From $y_{\text{A},0}$, we obtain $\rho_0$ and $\partial_{y_\text{A}}\rho$, $\sigma_0$ and $\partial_{y_\text{A}}\sigma$, as well as $\mu_0$, $D_{0}$ and $c_{\text{A},0}^{\text{eq}}$ from the composition-dependent properties of binary glycerol-water mixtures. This allows to calculate the normalized evaporation-induced composition gradient $\mbox{\it Ev}_y$ and the characteristic numbers $\mbox{\it Ra}$ and $\mbox{\it Ma}$. On the basis of these numbers and the contact angle, we solve the simplified quasi-stationary model and re-dimensionalise the resulting velocity and composition field as well as the evaporation rate using the scales \eqref{eq:simpmodel:scales}. This procedure allows to compare the full unsteady evolution of the droplet with the corresponding predictions at each instant by the simplified quasi-stationary model.

\begin{figure}\centering\includegraphics[width=1\textwidth]{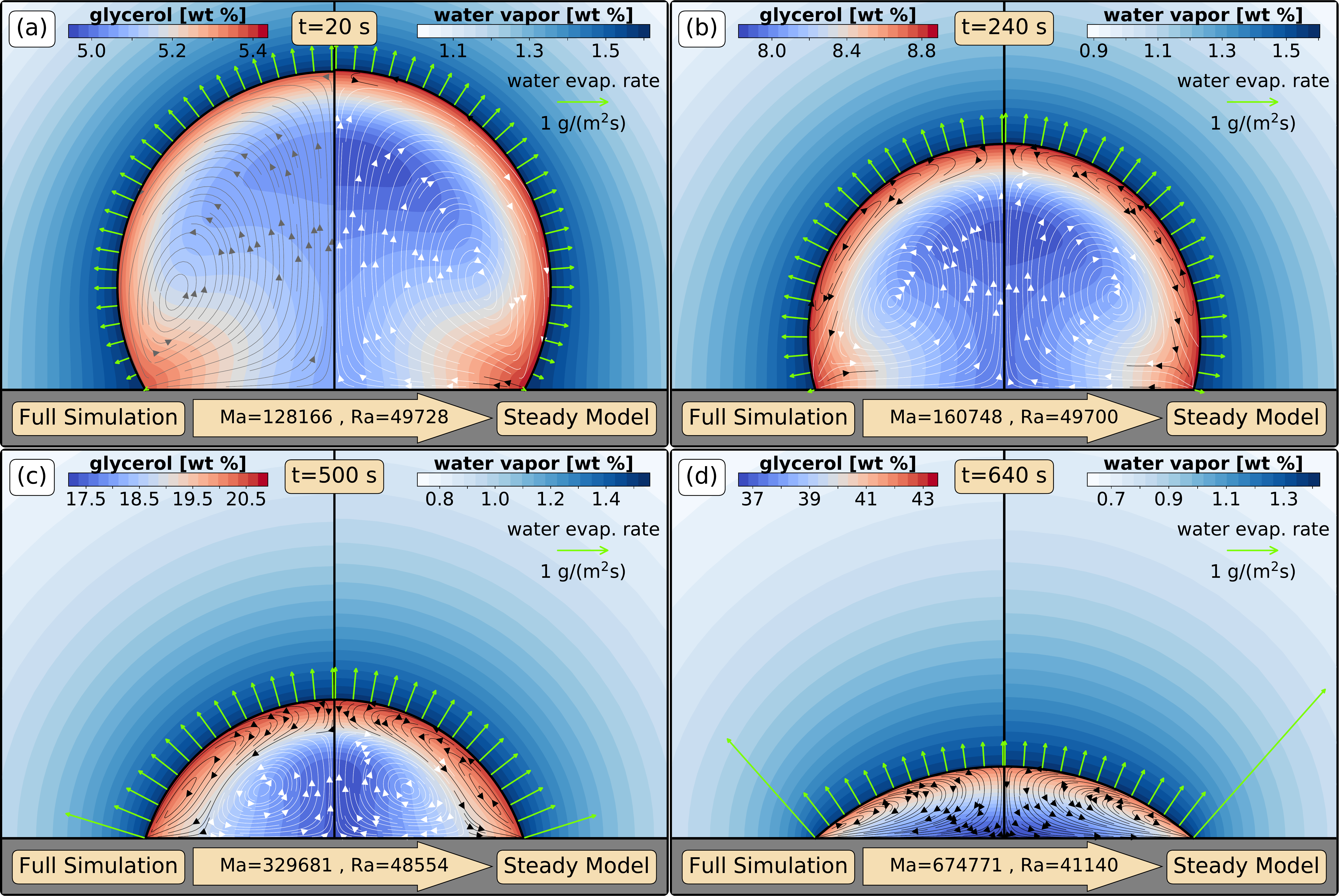}
\caption{Comparison of the full simulation (left) from \figurename~\ref{fig:motiv:example} and the corresponding result predicted by the quasi-stationary model (right) at different times. The colour-code inside shows the glycerol concentration, whereas the streamlines indicate the velocity field. In the gas phase, the water vapour and the corresponding evaporation rate is depicted. (a) Initially, the full simulation has not yet attained the quasi-stationary limit, so that the intensity of the composition deviation is overpredicted in the quasi-stationary model. In (b-d), the quasi-stationary model predicts the result of the full simulation up to a deviation that can be barely seen by eye. See supplementary movie 3 for the comparison between full simulation and quasi-stationary model over the entire simulation time. }
\label{fig:validation:numcompare}
\end{figure}

The results are depicted for several instants in \figurename~\ref{fig:validation:numcompare}, where the full simulation is shown on the left and the corresponding prediction of the quasi-stationary model is depicted on the right. Initially, i.e. in \figurename~\ref{fig:validation:numcompare}(a), the full simulation has not attained the quasi-stationary limit. Hence, the quasi-stationary model slightly overpredicts the composition variations, i.e. it shows more glycerol (red) near the interface and more water (blue) in the bulk. Therefore, the flow field also slightly differs, i.e. the transient full simulation shows a single vortex, whereas the quasi-stationary model predicts the presence of a small counter-rotating vortex near the apex. Furthermore, a very gentle deviation in the spherical cap shape due to the gravitational effect in the full simulation can be seen at the apex as well (regime \textsl{Ra reverses Ma}). At later time steps, i.e. in \figurename~\ref{fig:validation:numcompare}(b-d), however, the flow and the composition predicted by the quasi-stationary model match the results of the full simulation almost perfectly, be it in terms of the composition field, the flow pattern, the shape or the evaporation rate. This result substantiates the fact that the capillary outwards flow, which has been disregarded in the quasi-stationary model, can indeed be neglected as long as there is a prominent recirculating flow due to Marangoni and/or Rayleigh convection.

\begin{figure}\centering\includegraphics[width=1\textwidth]{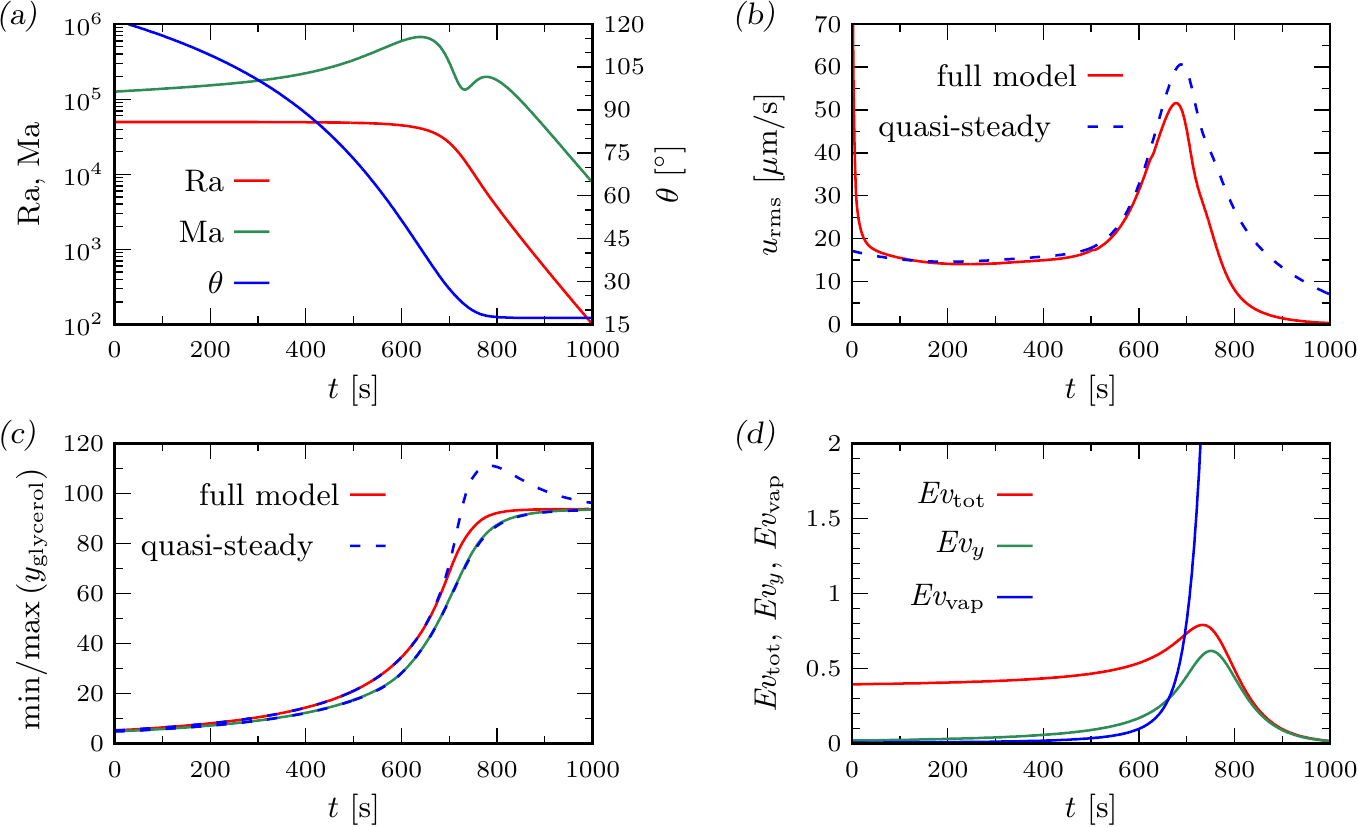}
\caption{Comparison of characteristic quantities of the full simulation of \figurename~\ref{fig:motiv:example} and the corresponding results predicted by the quasi-stationary model. (a) The three input parameters for the quasi-stationary model $\mbox{\it Ra}$, $\mbox{\it Ma}$ and $\theta$ extracted from the full simulation. (b) Comparison of the rms velocity. (c) Comparison of the maximum and minimum glycerol content inside the droplet. (d) Evaporation numbers extracted from the full simulation. }
\label{fig:validation:CompareQuants}
\end{figure}

To assess the quality of the quasi-stationary model in more detail, we have extracted some characteristic quantities of both simulations, i.e. from the full simulation of \figurename~\ref{fig:motiv:example} and the corresponding quasi-stationary limit at each instant. In \figurename~\ref{fig:validation:CompareQuants}(a), the time evolution of the three key parameters, namely the Rayleigh number $\mbox{\it Ra}$, the Marangoni number $\mbox{\it Ma}$ and the contact angle $\theta$ is shown. These numbers were used as input for the quasi-stationary model.  The rms of the velocity inside the droplet is depicted in \figurename~\ref{fig:validation:CompareQuants}(b). Again one can see an initial disagreement due to the fact that the full simulation has not yet attained its quasi-stationary limit. After that, i.e. after about $\SI{50}{\second}$ to $\SI{100}{\second}$, the rms-velocity is well predicted until it shows again a disagreement towards the end of the drying time. The reason for the overpredicted velocity in the quasi-stationary model can be found in \figurename~\ref{fig:validation:CompareQuants}(c), where the minimum and maximum glycerol concentration in the droplet according to both simulations are plotted against time. While it shows good agreement in the main part of the drying, the quasi-stationary model shows an enhanced maximum glycerol concentration towards the end of the drying, i.e. when almost only glycerol is left in the droplet. In fact, the glycerol concentration predicted by the quasi-stationary model even exceeds the physically realistic threshold of $\SI{100}{\percent}$. Obviously, this overprediction of the composition differences explains the elevated prediction of the rms velocity. The reason of the overpredicted composition difference can finally be seen in \figurename~\ref{fig:validation:CompareQuants}(d), where the evaporation numbers are depicted. When the droplet almost entirely consists of glycerol, the evaporation number $\mbox{\it Ev}_{\text{vap}}$, quantifying the reduction of the water vapour pressure for vanishing water at the interface on the evaporation dynamics (cf. Eq. \eqref{eq:simpmodel:evy}), becomes very large ($\mbox{\it Ev}_{\text{vap}}\to 18$ at $t=\SI{1000}{\second}$). This effect is not considered in the quasi-stationary model, since it assumes the averaged composition $y_{\text{A},0}$ to predict the vapour-liquid equilibrium, not the local composition at the interface. Thereby, the amount of water vapour is strongly overestimated which results in a high evaporation rate and thereby in an unrealistically high composition difference. Obviously, the quasi-stationary model loses validity when $\mbox{\it Ev}_{\text{vap}}$ becomes too large, meaning that the dependence of the vapour-liquid equilibrium on the local interface composition cannot be neglected any more. For a more detailed model, this effect can easily be incorporated into the quasi-stationary model, but it would introduce a fourth parameter besides $\mbox{\it Ma}$, $\mbox{\it Ra}$ and $\theta$ into the set of equations, which is beyond the scope of this article.
\section{Conclusion}
\label{sec:conclusion:conclusion}
During the evaporation of a binary droplet, multiple flow scenarios can be found, which is a result of an interplay of differences in the volatilities, mass densities and surface tensions of the two constituents. The difference in the volatilities induces compositional gradients in the bulk and also, due to the in general non-homogeneous evaporation rate, along the interface. Due to the composition-dependent mass density and surface tension, natural convection and Marangoni flow can set in, leading to a recirculating flow in the droplet that is usually much faster than the typical capillary outwards flow towards the contact line, which can be seen in pure droplets and leads to the coffee-stain effect in particle-laden droplets.

Based on justified assumptions, we simplified the full model equations to a quasi-stationary model that only requires three parameters, namely the contact angle, the Rayleigh and the Marangoni number. Both, the Rayleigh and Marangoni number, linearly scale with a non-dimensional evaporation number, $\mbox{\it Ev}_y$, which is a measure for the induced composition gradient by the preferential evaporation of one or the other component. By numerically solving for stationary solutions of the simplified model, we have explored the phase space in terms of these three quantities. The obtained phase diagrams allow for the prediction of the flow types in sessile and pendant binary droplets, with contact angles below and above \SI{90}{\degree}. 

We found in total five different flow patterns: If one of both mechanisms, i.e. either natural convection or Marangoni flow, gets sufficiently strong as compared to the other one, it can dominate and control the flow direction in the entire droplet. This scenario can usually be seen in the case when the corresponding number, namely the Rayleigh or Marangoni number, respectively, is much larger than the other. In these cases, a single vortex can be seen in the droplet. If both mechanisms drive the flow into a different direction and are comparably strong in terms of their non-dimensional numbers, one can find two vortices, one in the bulk driven by natural convection, and a counter-rotating vortex at the interface due to Marangoni flow.
The fourth flow type is the case, when both mechanisms act in the same direction, so that one cannot distinguish the particular cause of the driving and only a single vortex is present. Remarkably, however, in particular regimes in the phase space, the Marangoni flow can be reversed due to the natural convection in the bulk, leading to the fifth solution, where again two vortices can be found. In this situation, the bulk flow driven by natural convection deforms the internal composition field so that the diffusion dynamics in the liquid are altered, which eventually reverses the composition gradient at the interface and hence the Marangoni flow.

To use the phase diagrams presented in this article, several requirements have to be fulfilled: First of all, the influence of thermal effects must be negligible compared to the solutal ones. The two liquids must be miscible and the droplet must not be too large, so that the capillary number and the Bond number are small in order to guarantee a spherical cap shape during the evaporation. Furthermore, the Reynolds number must be small and the spatial variations in the composition must be small enough in order to allow for a first order Taylor expansion of the composition-dependent liquid properties according to \eqref{eq:simpmodel:compoexpand}. Also the requirements for the Bousinessq approximation must hold. The Marangoni number, as defined in \eqref{eq:simpmodel:MaNoRaNo}, must be positive, i.e. evaporation leads to an overall decrease of the surface tension, so that Marangoni-unstable chaotic flow can be excluded and the recirculating flow must be sufficiently faster than the movement of the interface. Finally, the influence of a change in the local composition on the vapour-liquid equilibrium may not be too strong, as it has been discussed on the basis on the evaporation number $\mbox{\it Ev}_{\text{vap}}$ describing the feedback of local composition changes on the evaporation rate in section \ref{sec:validation:validation}.

If all these requirements are fulfilled and the composition-dependence of the required physical properties are known, the phase diagrams of this article allow for a prediction of the qualitative flow pattern in an evaporating binary droplet, probably with exception of a short initial transient phase.

The method described in this article can be directly transferred to thermally driven Marangoni flow and natural convection in a pure droplet. Instead of a convection-diffusion equation for one component, one would have to consider the convection-diffusion equation for the temperature field. The boundary conditions will be different, e.g. a Dirichlet boundary condition of constant temperature at a highly conducting substrate and non-dimensional evaporative cooling instead of the number $\mbox{\it Ev}_y$, but the methodological principle can remain the same. 
Also a generalization to negative Marangoni numbers could be interesting, but it would require the consideration of the problem in three dimensions. This would allow to predict axial symmetry breaking and also bifurcations into chaotic Marangoni flow regimes by performing a linear stability analysis of the quasi-stationary solutions. 

\begin{acknowledgments}
This work is part of an Industrial Partnership Programme (IPP) of the Netherlands Organization for Scientific Research (NWO). This research programme is co-financed by Canon Production Printing Holding B.V., University of Twente and Eindhoven University of Technology. DL gratefully acknowledges support by his ERC-Advanced Grant DDD (project number 740479). 
\end{acknowledgments}

\section*{Declaration of Interests}
The authors report no conflict of interest.

\section*{Appendix A. Comparison with experiments and relation to the Grashof number}
\label{sec:appendix:appendix}
Since even the detailed full model is subject to some assumptions, e.g. the diffusion-limited vapour transport and the disregard of thermal effects, we also performed experiments on various sessile and pendant binary droplets with different volumes and contact angles. The details of the experimental setup are described by \citet{Li2019}.
Here, we are more interested in a qualitative agreement, i.e. whether the flow direction is dominated by natural convection or not. 

Simultaneously, we address the Grashof number (also known as Archimedes number) in the following. This number, defined as 
\begin{equation}
\mbox{\it Gr}=gh^3\rho_0(\rho_\text{A,pure}-\rho_\text{B,pure})/\mu^2\,,
\label{eq:appendix:grass}
\end{equation}
with $h$ being the height of the droplet and $\mu$ the averaged viscosity of both liquids, was used in our previous publication \citep{Li2019} as an indicator whether the flow in the droplet is dominated by natural convection ($\mbox{\it Gr}\gg 1$) or not ($\mbox{\it Gr}\ll 1$). Compared to the non-dimensional numbers presented in this manuscript, i.e. the evaporation number $\mbox{\it Ev}_y$ and the Rayleigh number $\mbox{\it Ra}$, this number is independent of the current droplet composition. Instead, it just takes the pure densities of both fluids and the averaged density and viscosity into account. This means that the Grashof number $\mbox{\it Gr}$ is easily accessible, whereas the non-dimensional numbers used throughout this manuscript require the knowledge of the instantaneous average composition and the full composition-dependence of all properties, which is not always possible in an experimental setup.

Therefore, we are interested to substantiate the argumentation by \citet{Li2019} that the much simpler Grashof number $\mbox{\it Gr}$ can be used as indicator whether to expect natural convection ($\mbox{\it Gr}\gg 1$) or not ($\mbox{\it Gr}\ll 1$).
To investigate the validity, we replace the Rayleigh number by the Grashof number in the following. If one assumes that $\partial_{y_\text{A}}\rho$ is independent of the composition, i.e. a linear dependence of the mass density on the mass fractions, one can obtain the Grashof number via the relation  
\begin{equation}
\mbox{\it Gr}=\frac{3}{\pi}\, \frac{1-\cos \theta}{2+\cos \theta}\,\frac{\mbox{\it Ra}}{\mbox{\it Ev}_y\mbox{\it Sc}}\,,
\label{eq:appendix:grashof}
\end{equation}
where the factor depending on the contact angle $\theta$ is a consequence of the different characteristic length scales, i.e. $\sqrt[3]{V}$ for $\mbox{\it Ra}$ and $h$ for $\mbox{\it Gr}$.
While the Schmidt number $\mbox{\it Sc}=\mu/(\rho D)$ for liquids is typically $\mbox{\it Sc}\sim\mathcal{O}(10^4-10^5)$, for moderately volatile liquids like e.g. water at typical ambient conditions, $\mbox{\it Ev}_y\sim\mathcal{O}(10^{-1}-10^0)$ holds. In order to obtain diagrams independent of these quantities, we set the factor $\mbox{\it Ev}_y\mbox{\it Sc}=1000$ in \eqref{eq:appendix:grashof} for the determination of the boundaries in the phase diagrams.

\begin{figure}\centering\includegraphics[width=1\textwidth]{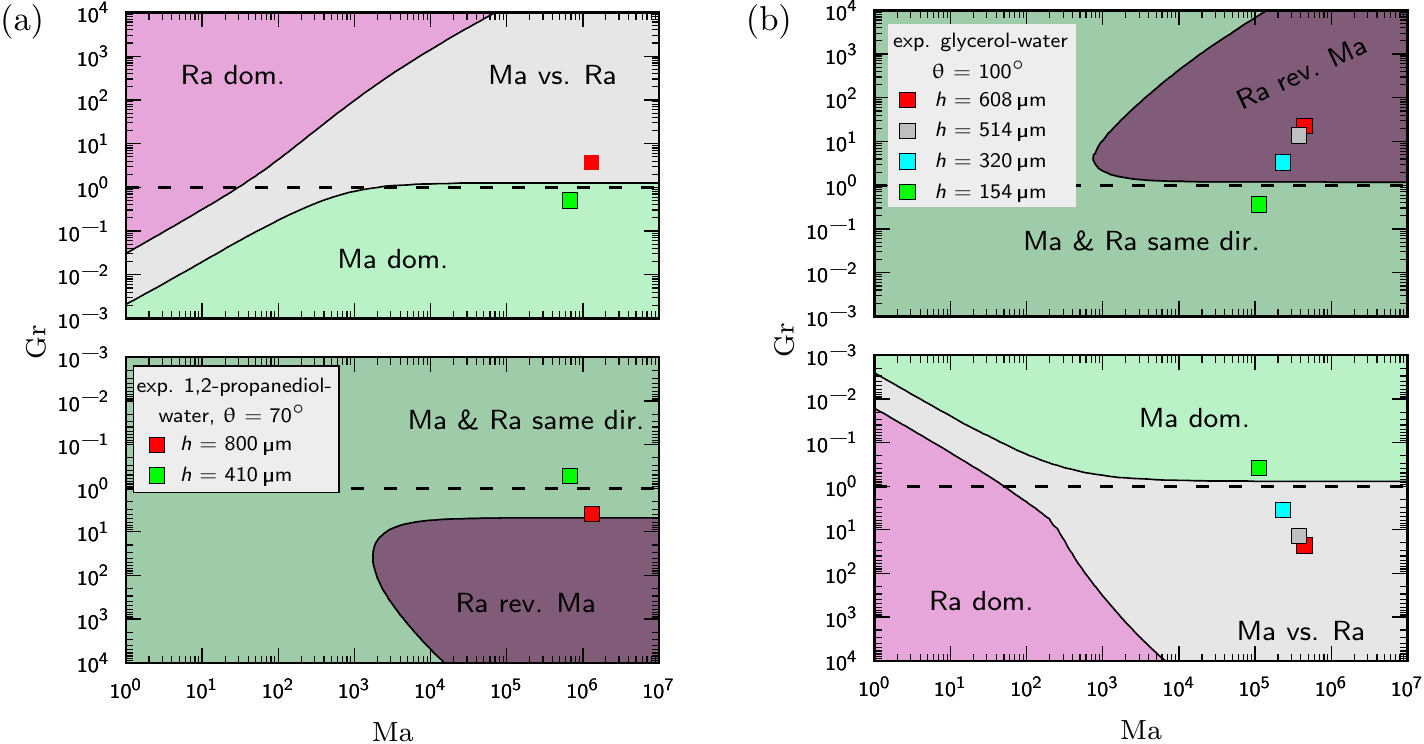}
\caption{Same as \figurename~\ref{fig:phasediag:MaRaLowCa} and \figurename~\ref{fig:phasediag:MaRaHighCa}, but expressed in terms of $\mbox{\it Gr}$ instead of $\mbox{\it Ra}$. Obviously, the onset of gravity-driven flow, even in presence of rather strong Marangoni driving happens close to $\mbox{\it Gr}=1$ (indicated by the grey line). Furthermore, experimental data of \citet{Li2019} is also indicated.}
\label{fig:appendix:ArMont}
\end{figure}

The phase diagrams rescaled to the Grashof number via this way are depicted in \figurename~\ref{fig:appendix:ArMont}.
One can infer from these diagrams that even in competition with a strong Marangoni effect, the onset of gravity-driven bulk flow happens approximately at a Grashof number of $\mbox{\it Gr}\sim\mathcal{O}(1)$ for a contact angle of $\theta=\SI{70}{\degree}$ in (a) and $\theta=\SI{100}{\degree}$ in (b). Furthermore, the experimental results of \citet{Li2019} are indicated as dots. The 1{,}2-propanediol-water droplets with a contact angle of $\theta=\SI{70}{\degree}$ discussed in \citet{Li2019} clearly show the effect of natural convection for an apex height of $h=\SI{800}{\micro\meter}$, whereas is was not visible for $h=\SI{410}{\micro\meter}$. This clearly coincides with the prediction of the phase diagram in \figurename~\ref{fig:appendix:ArMont}(a). The experiments on glycerol-water droplets with $\theta=\SI{100}{\degree}$, as discussed in the supplementary information of \citet{Li2019}, reveal an absence of observable natural convection for $h=\SI{154}{\micro\meter}$, whereas the presence of natural convection was found at heights $h\geq\SI{320}{\micro\meter}$, with increasing velocity for elevated heights. Also this can be inferred from the $\mbox{\it Ma}$-$\mbox{\it Gr}$-diagram depicted in \figurename~\ref{fig:appendix:ArMont}(b). Thus we conclude that the Grashof number $\mbox{\it Gr}$ is indeed an indicator for the presence or absence of decisive natural convection in a binary droplet. The $\mbox{\it Ma}$-$\mbox{\it Ra}$-diagrams presented in this manuscript, however, provide a much more detailed prediction of the possible flow scenarios.

\bibliographystyle{jfm}

\bibliography{refs}

\end{document}